\documentclass[]{aa}  
\usepackage{graphicx}
\usepackage{txfonts}
\usepackage{mathabx}
\usepackage{caption}
\usepackage{subcaption}
\usepackage{natbib}
\usepackage{placeins}
\usepackage{textcmds}
\begin{document} 
\newcommand{\ME}{\(\text{M}_\Earth\)\,}
\newcommand{\ca}{$\sim$}

   \title{Feasibility of interferometric observations and characterization of planet-induced structures at sub au to au scales in protoplanetary disks}
    \titlerunning{Observation and characterization of disk substructures at sub au to au scales}
   \author{L. Hildebrandt
          \inst{1}
          ,
          A. Krieger
          \inst{1}
          ,
          H. Klahr
          \inst{2}          
          ,
          J. Kobus
          \inst{1}
          ,  
          A. Bensberg
          \inst{1}
          , and
          S. Wolf
          \inst{1}
          }
    \authorrunning{L. Hildebrandt et. al.}
   \institute{Institute of Theoretical Physics and Astrophysics, University of Kiel, Leibnizstraße 15, 24118 Kiel, Germany\\         \email{lhildebrandt@astrophysik.uni-kiel.de}\label{inst1}  \and
              Max-Planck-Institut für Astronomie, Königstuhl 17, 69117 Heidelberg, Germany\label{inst2}
             }

  \abstract 
   {
   Interferometric observations of protoplanetary disks by VLTI and ALMA have greatly improved our understanding of the detailed structure of these planetary birthplaces.
   These observations have revealed a variety of large-scale disk substructures, including rings, gaps, and spirals, spanning tens to hundreds of au, supporting the predictions of planet formation models.
   Recent instruments, such as MATISSE at the VLTI, allow one to resolve and investigate the inner few au of protoplanetary disks in nearby star formation regions, shedding light on the traces of planet formation and evolution at these small scales.
   }
   {
   The aim of this work is to assess the feasibility of interferometric observations of small-scale planet-induced substructures in protoplanetary disks in nearby star-forming regions.
   We aim to characterize these substructures in multi-wavelength and multi-epoch observations and subsequently differentiate between simulation parameters.
    }
  {
  On the basis of 3D hydrodynamic simulations of embedded planetary companions and subsequent 3D Monte Carlo radiative transfer simulations, we calculated and analyzed interferometric observables, assuming observations with the VLTI in the K, L, M, and N bands. 
   }
   {
   The hydrodynamic simulations exhibit mass-dependent planet-induced density waves that create observable substructures, most notably for the considered case of a 300 \ME planet. These substructures share similarities with observed large-scale structures and feature a prominent accretion region around the embedded planet. The visibilities show a detectable variability for multi-epoch VLTI/GRAVITY and VLTI/MATISSE observations, caused by the orbital motion of the planet, that are distinguishable from other sources of variability due to their unique combination of timescale and amplitude. Additionally, the non-uniform change of the visibilities at different baselines can be used to identify asymmetric structures. Furthermore, we show that multi-wavelength observations provide an approach to identify the fainter substructures and the signal of the accretion region.
    }
   {}

   \maketitle
   
\section{Introduction}\label{chap:introduction}

Protoplanetary disks (PPDs), the birthplaces of planetary systems, are dynamic environments in which planets form and evolve. The embedded planets and their influence on the structure and evolution of the disk determine the final architecture of planetary systems via a multitude of processes \citep[e.g.,][]{safronov1972,williams2011}.
Current models on planet formation are incomplete in some areas, such as the dust growth process where models show growth barriers that have to be overcome \citep[e.g.,][]{seizinger2013,testi2014,drazkowska2023}.
Although a direct resolved observation of an embedded planet at this stage of its evolution is currently not feasible due to the small angular size and obscuring dust and gas, there are two effects caused by a planet at this early stage that are potentially easier to observe.
One effect is the accretion onto the planet, where hot dust in the vicinity causes a local brightness increase due to the thermal reemission by the dust. 
The second effect is the characteristic disk density structures resulting from the planet-disk interaction (e.g., \citealt{andrews2020}), which are part of recent studies (e.g., \citealt{Carrera_2021}). 

While early hydrodynamic models predicted the existence of disk substructures, only the first high-resolution observations of PPDs gave way to advanced planet formation models with respect to the observed effects of local disk architecture and initial conditions.
The observations of the Atacama Large Millimeter/submillimeter Array \citep[ALMA; ][]{kurz2002} of PPDs at scales of dozens to hundreds of au showed a general feasibility of observing disk substructures to provide insight into the disk formation \citep{hsieh2024} and early phases of planet formation \citep{Ohashi_2023}.
Thanks to the increasing availability of high-resolution instruments, a growing number of PPDs have been observed with sufficient resolution to detect disk substructures \citep[see, e.g.,][]{boekel2017, momose2015, stolker2016, bohn2019, benisty2017}. 
The angular resolution required to observe the inner regions of PPDs in nearby star formation regions ($\lessapprox10$ milliarcseconds, corresponding to au to sub au scales at typical distances of 140 pc) can only be achieved with interferometric instruments, whereas the dust of the PPD is preferably observed in the infrared. 
For two decades, instruments such as the MID-infrared Interferometric instrument \citep[MIDI;][]{leinert2003} in the N band and the Astronomical Multi-BEam combineR \citep[AMBER;][]{petrov2007} in the J, H, and K band provided observations and insight into disk brightness distributions with a resolution of a few au \citep[see, e.g.,][]{boekel2006}.

Previously observed large-scale substructures have been categorized into morphological classes such as rings, gaps, spirals, crescents, and shadows. A detailed overview can be found in \citet{bae2023}. Both the presence of multiple substructures in a single disk and wavelength-dependent substructures have been observed \citep{dong2018}. 
Early investigations by, for instance, \citet{wolf2005} already indicated the possibility for a successful detection of induced substructures from Jupiter-mass planets for nearby systems, closer than 100 pc, and planets at distances of 5 au from the star with ALMA.
Only recent interferometric instruments at the Very Large Telescope Interferometer (VLTI), namely GRAVITY, its upgrade GRAVITY+, and the Multi-AperTure mid-Infrared SpectroScopic Experiment (MATISSE), are able to resolve the central regions of PPDs and allow observations of the inner regions at smaller scales \citep[][]{gravity2017, gravity2022, lopez2022}. 

The innermost regions of a stellar system on scales of a few au down to sub au separation from the star are known to host hot Jupiters \citep{mayor1995}. 
Since the formation process of these bodies is not yet fully explained, there exist different theories of where and how hot Jupiters form \citep[see, e.g.,][]{lin1996}, with one of them being an in situ explanation \citep{batygin2016}.
Insights into the local disk environment of these regions could be used to verify the different planet formation theories.
In addition, it is of particular interest to understand the formation process of terrestrial and potentially habitable planets, which requires detailed information of the initial conditions at terrestrial scales of \ca1 au orbital radius.
It is therefor essential to observe the inner few au of PPDs to investigate potentially embedded planets or disk structures, as they could provide new constraints on existing planet formation theories.\\
This work focuses on the possible traces of embedded planets at 1 au orbital distance, particularly density wave-induced disk substructures; their potential observability in the near and mid-infrared; and whether a distinction from other sources of variability in the observations is possible.

For observations in the L, M, and N bands, the instrument of choice is MATISSE of the VLTI \citep{lopez2022} consisting of the small movable Auxiliary Telescopes (ATs) and the larger, more sensitive Unit Telescopes (UTs).
For the objects studied in this work, potential observations require the use of the UTs, because of the faintness of the PPDs that prohibits the use of the ATs.
For example, for N band measurements, a total flux of 17 Jy is required \citep{petrov2020}, whereas the simulated disks in this study have a total flux of less than 2 Jy.
The simulated K band observations consider the GRAVITY instrument at the VLTI, which provides a resolution of up to 4 milliarcseconds, corresponding to structures of \ca 0.56 au size in our simulations and requires a target flux of at least 0.0667 Jy \citep{gravity2017}, which is surpassed by the simulated K band flux in this work.

While a possible approach to observing small-scale substructures lies in kinematic observations of density perturbations, caused by planet-disk interactions \citep{fasano2024}, difficulties arise with interfering effects, such as disk wind fluctuations in the inner 1.5 au of the disk \citep{chambers2024}. 
These effects may limit the accuracy of observations of substructures. An alternative approach uses the photometric variability of these young systems \citep[see, e.g.,][]{manick2024}, which is a common and characteristic feature \citep[see, e.g.,][]{kobus2020, {bensberg2023}}. 
The photometric and spectroscopic variability of these systems was found to be on timescales between days and years, as shown by observations of PDS 70 by \citet{gaidos2024}. 
Any planet-induced interferometric variability will have to rely on the movement of asymmetric substructures around the star, over timescales corresponding to the orbital distance of the planet or the substructures.
For an object with a revolving environment, such as PPDs, two observations at sufficiently spaced points in time measure different points in the uv-plane, since a rotation in the image plane also rotates its 2D Fourier transform.
However, it has to be considered that multiple sources of variability exist for these young systems, so further differentiation between interferometric variability from asymmetric disk substructures and other sources has to be made. 

In Sect. \ref{chap:methods} we describe the hydrodynamics simulations and the Monte Carlo radiative transfer code we use in this work. 
The subsequent results of our simulated PPDs are shown in Sect. \ref{chap:results}, with the density maps in Sect. \ref{chap:density}, flux maps in Sect. \ref{chap:flux}, visibilities in Sect. \ref{chap:visibility}, and closure phases in Sect. \ref{chap:closphase}.
We further compare the results with existing observations in Sect. \ref{chap:submorph}, assess the observability of the substructures in Sect. \ref{chap:observ}, provide a modeling approach for fainter substructures in Sect. \ref{chap:modeling}, and discuss approaches to differentiate the detectable variability from other, non planet-induced sources in Sect. \ref{chap:differentiation}.

\section{Methods}\label{chap:methods}

In our study, we utilize the TRAMP hydrodynamics code, which is detailed in \citet{1999ApJ...514..325K}. This code has been augmented by incorporating a radiation hydrodynamics scheme, as explored in \citet{kuiper2010}. Furthermore, the code has undergone additional refinements, as elaborated in \citet{krieger_klahr24}.

\begin{figure}
    \centering
    \includegraphics[width=.9\linewidth]{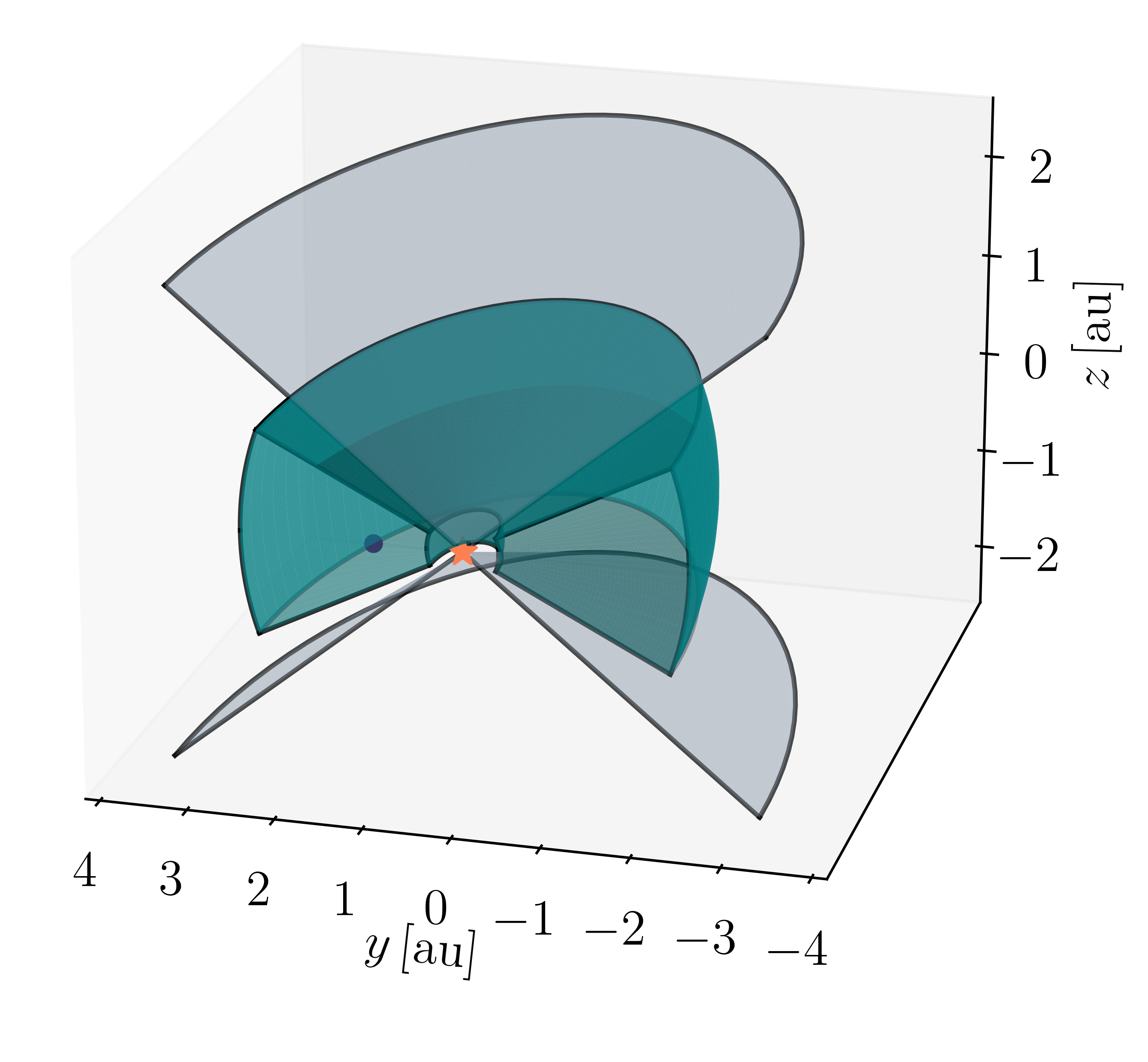}
    \caption{Different regions of the computational grid. The high-resolution grid (green) is embedded in the low-resolution background grid (between the gray sections). The star (orange) is shown at the center of the grid, while the embedded planet (blue) is located at an orbital distance of 1 au. The grid is rotationally symmetric in the xy-plane.}
    \label{fig:grid_example}
\end{figure}

We modeled a typical protoplanetary disk with an accretion rate of  $4.36 \times 10^{-9}\,{\rm M}_\odot {\rm yr}^{-1}$ around a $0.5\,{\rm M}_\odot$ star with a radius of $R_* = 2.5\,{\rm R}_\odot$ and an effective temperature of $T_* = 4000\,$K, which results in a luminosity of $L=1.43\,{\rm L}_\odot$. 
In our simulations, we employed an alpha viscosity parameter of $\alpha = 3 \times 10^{-3}$.
This parameter is the ratio of viscous stress to gas pressure in the disk.
It causes faster accretion by more efficient angular momentum transport for higher values and was first introduced by \citet{shakura1973}.
The disk in our model has a total mass (gas and dust) of $1\%$ of the solar mass. 
We introduced a planet embedded at a radius of 1 au within this disk, which interacts with the surrounding material, leading to the formation of various substructures in the disk due to its gravitational influence.
The hydrodynamic model was calculated based on a 3D high-resolution grid with a resolution of 101 radial cells for radii between 0.4 and 2.5 au, 339 cells along the full azimuthal range, and 53 cells for the polar angle between 45 degrees to either side of the disk plane.
This high-resolution grid was further embedded in a lower resolution axisymmetric background disk model using 200 radial and 100 polar cells spanning a radial range of from an inner disk radius of 0.1 au to an outer radius of 300 au and a polar range of 60 degrees.
A schematic overview of this computational grid can be seen in Fig. \ref{fig:grid_example}. 

For our analysis, we examined snapshots from these hydrodynamical (HD) simulations to study the presence and observability of the resulting disk substructures in detail. 
The results of our HD simulations were post-processed with the Monte Carlo radiative transfer (MCRT) code Mol3D \citep{ober2015}, which allowed us to derive a temperature distribution of the dust under the assumption of a radiative equilibrium. 
The simulated radiation sources emit photon packages that traverse the model space on randomly determined paths. 
The corresponding random variables follow probability density distributions, which depend on the wavelength-dependent optical properties of the medium. 
Interactions between photon packages and the medium, which include scattering, absorption, and reemission, are based on Mie theory. 
Photon packages are characterized by their wavelength, propagation direction, polarization state, and carried energy. 
As they traverse the model space, a portion of their energy is deposited into the medium, which results in a local temperature increase. 
Moreover, the photon packages that leave the systems are used to construct ideal flux maps.

The considered simulated radiation sources include the central star, a protoplanetary disk, as well as an embedded accreting planet. 
During the HD simulations, the planet accretes material, which gives rise to the emission of accretion luminosity. In our MCRT simulations, we use the derived accretion luminosity, which is assumed to be emitted from the planetary surface following a black body spectrum.
For this, we chose a planetary radius of 1 Jupiter radius, similar to \citet{krieger_klahr24}, due to a lack of observed radii of embedded protoplanets. 

In order to determine a realistic temperature distribution, it is crucial to ensure that the dust does not exceed its sublimation temperature. Therefore, all dust has been removed from cells with temperatures exceeding the dust sublimation temperature. Radiation originating from viscous dissipation is assumed to be fully thermalized and emitted by the dust phase of the protoplanetary disk, following their particular temperature and wavelength-dependent spectra. 

The dust grains are assumed to be spherical with radii $a$ ranging from $5\,$nm to $250\,$nm. Their sizes follow a grain size distribution d$n \sim a^{q_g} \text{d}a$ with exponent $q_g=-3.5$ \citep{1977ApJ...217..425M}. 
While larger dust grains with radii greater than $250\,$nm are expected to occur in these disks, the study of their influence on both the hydrodynamic and radiative transfer simulations was beyond the scope of this work and motivates a follow up investigation.
Every dust grain is composed of $f_{\rm Si}=62.5\,\%$ astronomical silicate and $f_\text{Gr}=37.5\,\%$ graphite using a $1/3\,$:$\,2/3$ approximation \citet[i.e., $f_{\parallel\text{-Gr}}\,$:$\,f_{\perp\text{-Gr}}$=1\,:\,2;][]{1993ApJ...414..632D} and a bulk density of $2.5\,\text{g}\,\text{cm}^{-3}$. 
In order to calculate the cross-sections under the assumption of Mie-theory, we applied the code miex \citep{2004CoPhC.162..113W} using a list of their corresponding wavelength-dependent refractive indices \citep{1984ApJ...285...89D,1993ApJ...402..441L,2001ApJ...548..296W} and finally averaged the resulting optical properties of the dust grains \citep{Wolf_2003}. The dust-to-gas ratio is set to a generic value of $1{:}100$ for all dust containing cells. 

In order to reduce the simulation time and computational cost, we applied different performance enhancing techniques. These include a method of locally divergence-free continuous absorption of photon packages \citep{1999A&A...344..282L}, an immediate photon package reemission scheme \citep{2001ApJ...554..615B}, as well as a method that utilizes a database of precalculated photon package paths \citep{2020A&A...635A.148K}.

Following the calculation of the dust temperature distribution, we derive realistic, wavelength-dependent flux maps. Besides the direct, although attenuated planetary and stellar radiation, the thermal reemission radiation of the dust (including dust self-scattering) and the scattered stellar and planetary radiation are considered. 
In order to further enhance the quality of our determined flux maps, we performed a dedicated monochromatic flux determination step for each considered observing wavelength for each considered source of radiation. 
In particular, we used each $10^7$ photon packages for the stellar and the planetary radiation. 
The flux contribution of the thermal dust emission was further subdivided into two steps. 
To account for the (direct) un-scattered part, we applied a non-probabilistic raytracer routine. 
This method has the advantage of providing a smooth and precise result that is not affected by noise, which is inherent to any Monte Carlo simulation. 
The contribution of self-scattered thermal dust emission, was then performed based on a MCRT simulation, where each cell emits a specific luminosity corresponding to its particular dust temperature and density.
In order to ensure energy conservation during these simulations, the individual energy carried by each simulated photon package depends on the cell it originates from, and is chosen to be less than but similar to the energy of a stellar photon package. 
The final flux map for each observing wavelength is obtained by adding up the derived flux maps of all radiation sources. 

In the following, we consider the observing wavelengths 2.12 $\mu$m  for the K band, 3.58 $\mu$m  for the L band, 4.99 $\mu m$ for the M band, and 10.06 $\mu$m for the N band, to simulate flux maps of three embedded planets with 1, 10, and 300 \ME. 
Eventually, a 2D Fourier transform of these flux maps is calculated with the Python library \texttt{galario} \citep{tazzari2018}, based on which the interferometrically observable quantities (visibilities and closure phases) are determined (see Sect. \ref{chap:visibility} and \ref{chap:closphase}).

Motivated by distances to nearby star-forming regions, such as Taurus-Auriga, we apply a distance of 140 pc in the following observational feasibility studies. 
At this distance, a structure of 1 au size in the disk, corresponds to an angular size of 7.14 milliarcseconds.
This resolution can be exceeded or at least approached by interferometric observations of both GRAVITY and MATISSE at the VLTI, with a resolution of 4 milliarcseconds in the K band \citep{gravity2017}, 5 milliarcseconds in the L band, \ca6 milliarcseconds in the M band, and 12.5 milliarcseconds in the N band \citep{lopez2022}. 
For the analysis presented in the following sections, the simulated PPDs were considered to be located at the celestial coordinates of HL Tau, which is a young stellar object in the above star forming region with a prominent protoplanetary disk, which has a declination of +18.23\textdegree.

\section{Results}\label{chap:results}
The following analysis of the influence of an embedded planet on the protoplanetary disk and its detectability is performed based on a combination of hydrodynamic and radiative transfer simulations, and their subsequent simulated visibilities and closure phases.
In Sect. \ref{chap:density} we present the results of the hydrodynamic simulations and describe the mass-dependent effect of the embedded planet. 
The resulting flux maps of the radiative transfer simulations and the wavelength dependency of the brightness contrast of the substructures is subsequently shown in Sect. \ref{chap:flux}.
An analysis of the visibilities and closure phases from these flux maps is presented in Sect. \ref{chap:visibility} and \ref{chap:closphase}, respectively.

\subsection{Simulated density maps}\label{chap:density}

\begin{figure}
     \centering
     \begin{subfigure}[b]{0.5\textwidth}
         \centering
         \includegraphics[width=\textwidth, trim={0 12 0 0}]{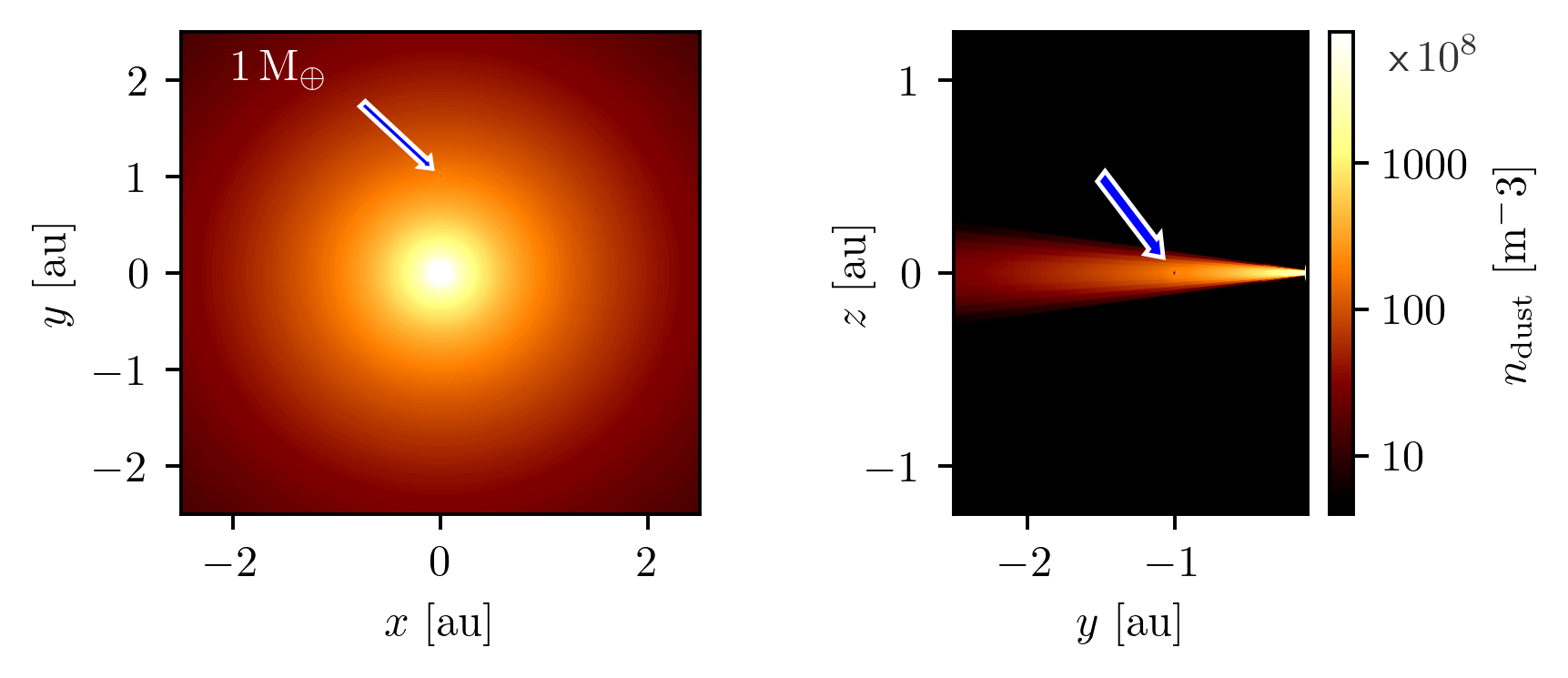}    
         \caption{$\text{M}_{\text{Planet}}$ = 1\,\ME } 
         \label{fig:dust1}
     \end{subfigure}
     \newline
     \vspace{0.125cm}
     \begin{subfigure}[b]{0.5\textwidth}
         \centering
         \includegraphics[width=\textwidth, trim={0 12 0 0}]{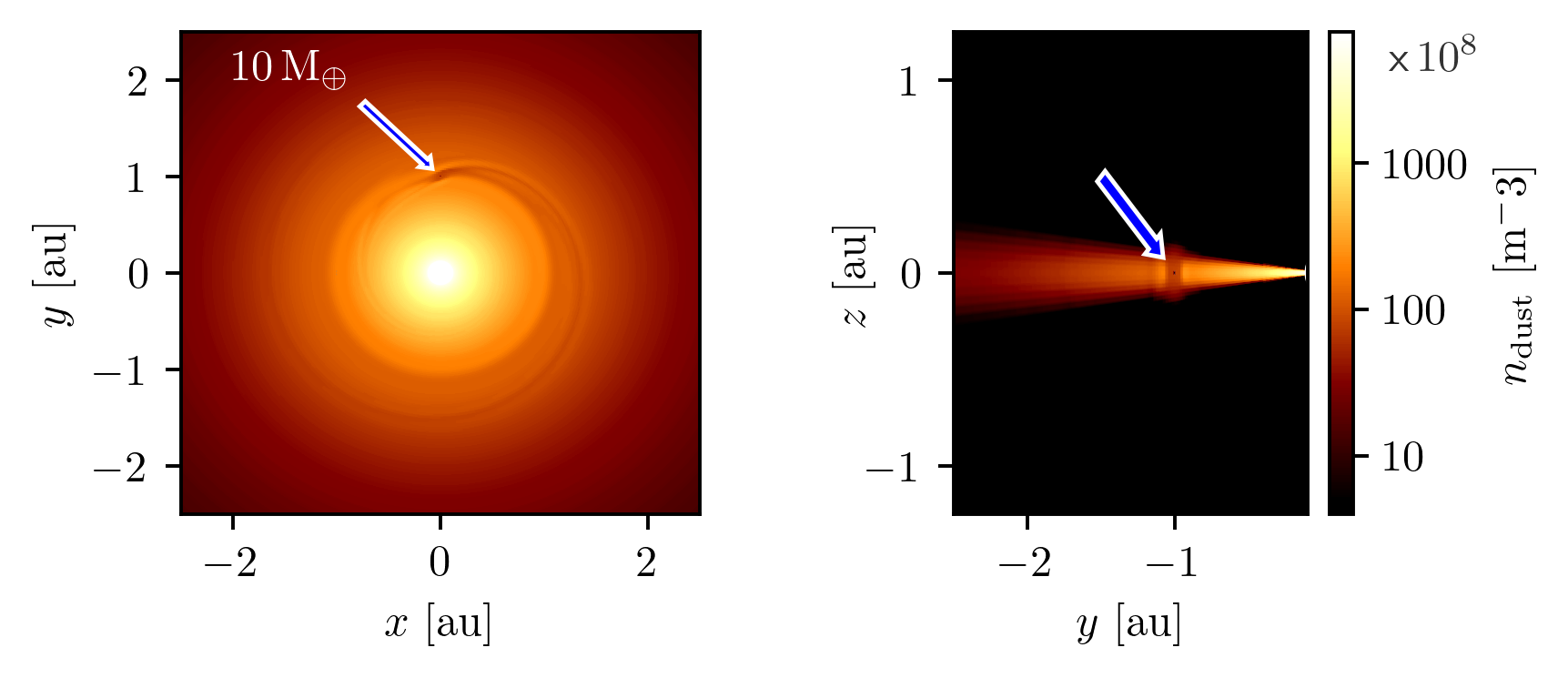}         
         \caption{$\text{M}_{\text{Planet}}$ = 10\,\ME } \label{fig:dust10}
     \end{subfigure}
     \newline
      \vspace{0.125cm}
     \begin{subfigure}[b]{0.5\textwidth}
         \centering
         \includegraphics[width=\textwidth, trim={0 12 0 0}]{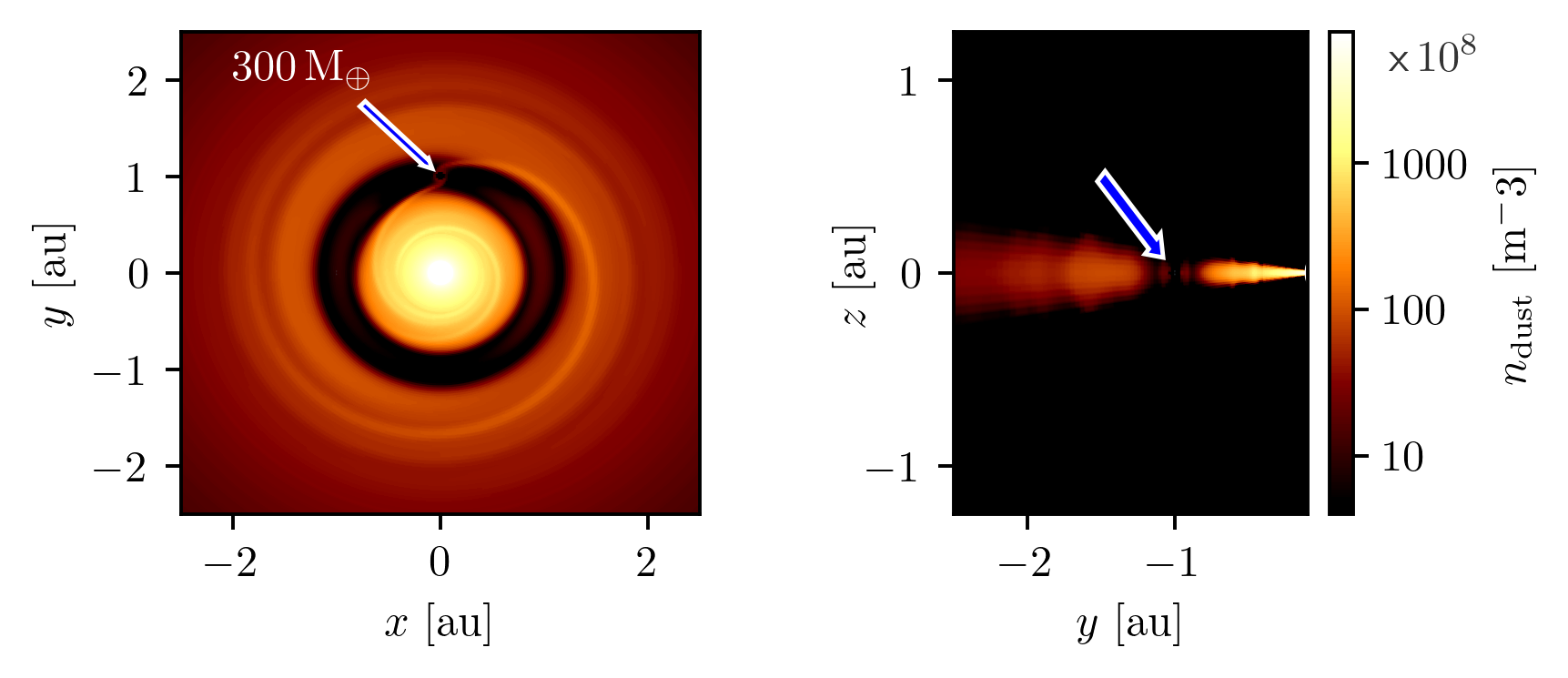}         
         \caption{$\text{M}_{\text{Planet}}$ = 300\,\ME } \label{fig:dust300}
     \end{subfigure}     
    \caption{Dust densities in the midplane (left) and perpendicular to the midplane across the position of the planet (right) in the hydrodynamic simulations for planet masses of 1\,\ME (top), 10\,\ME (middle), and 300\,\ME (bottom). The blue arrows indicate the position of the planet.} 
    \label{fig:dustdensities}
\end{figure}

Based on the aforementioned hydrodynamic simulations (see Sect. \ref{chap:methods}), the effect of the planet mass on PPDs can be seen in Fig. \ref{fig:dustdensities}. The simulated dust densities in the midplane are shown up to a radius of 2.5 au (left images) and radial profiles across the position of the planet up to 1.5 au in vertical direction (right images).
The planet is positioned at a radial distance of 1 au north from the central star in the images showing the density and flux maps (blue arrow).
Its influence in the dust densities increases with planet mass, from a localized density minimum in the vicinity of the 1 \ME  planet in the top image (Fig. \ref{fig:dust1}), to a warping in the dust densities and a planet-centered bulge in the radial cut for 10 \ME (Fig. \ref{fig:dust10}), to a significant gap for both cuts in the case of a 300 \ME planet (Fig. \ref{fig:dust300}). 

For the 10 \ME planet in Fig. \ref{fig:dust10}, a spiral-like feature emanates from a ring of higher dust density at the position of the planet. Similarly, a faint spiral arm is arcing inward from the planet toward the center region. Both spiral arms represent an increased dust density, compared to their respective background levels.

The further increased planet mass in the simulation of a 300 \ME planet (see Fig. \ref{fig:dust300}) features the same behavior with more pronounced spiral arms, both inward and outward from the planetary orbit. Additionally, an approximately 0.3 au wide gap exists at the orbital radius, representing dust densities that are diminished by more than one order of magnitude.
This indicates that the embedded planet is massive enough for sufficiently strong hydrodynamic disk interactions to form a gap along its entire orbital path and potentially create observable substructures on the disk surface.

\subsection{Simulated flux maps}\label{chap:flux}
In order to assess the observability of potential disk substructures, we first derive its corresponding wavelength-dependent brightness distribution.
For this work, we use the 3D MCRT code Mol3D, as described in Section \ref{chap:methods}.
The used wavelength bands correspond to those observed with GRAVITY and MATISSE at the VLTI.

The observability of substructures is determined not only by their relative brightness to the star, but also by the brightness contrast between each characteristic structure and the local disk background.
For this reason, a flux map at a certain wavelength may show a higher overall disk flux, but less observable substructures, due to a smaller contrast, while another wavelength may show a fainter disk, with better observable substructures.

 \begin{figure*}[p]
     \centering
     \includegraphics[width=\textwidth]{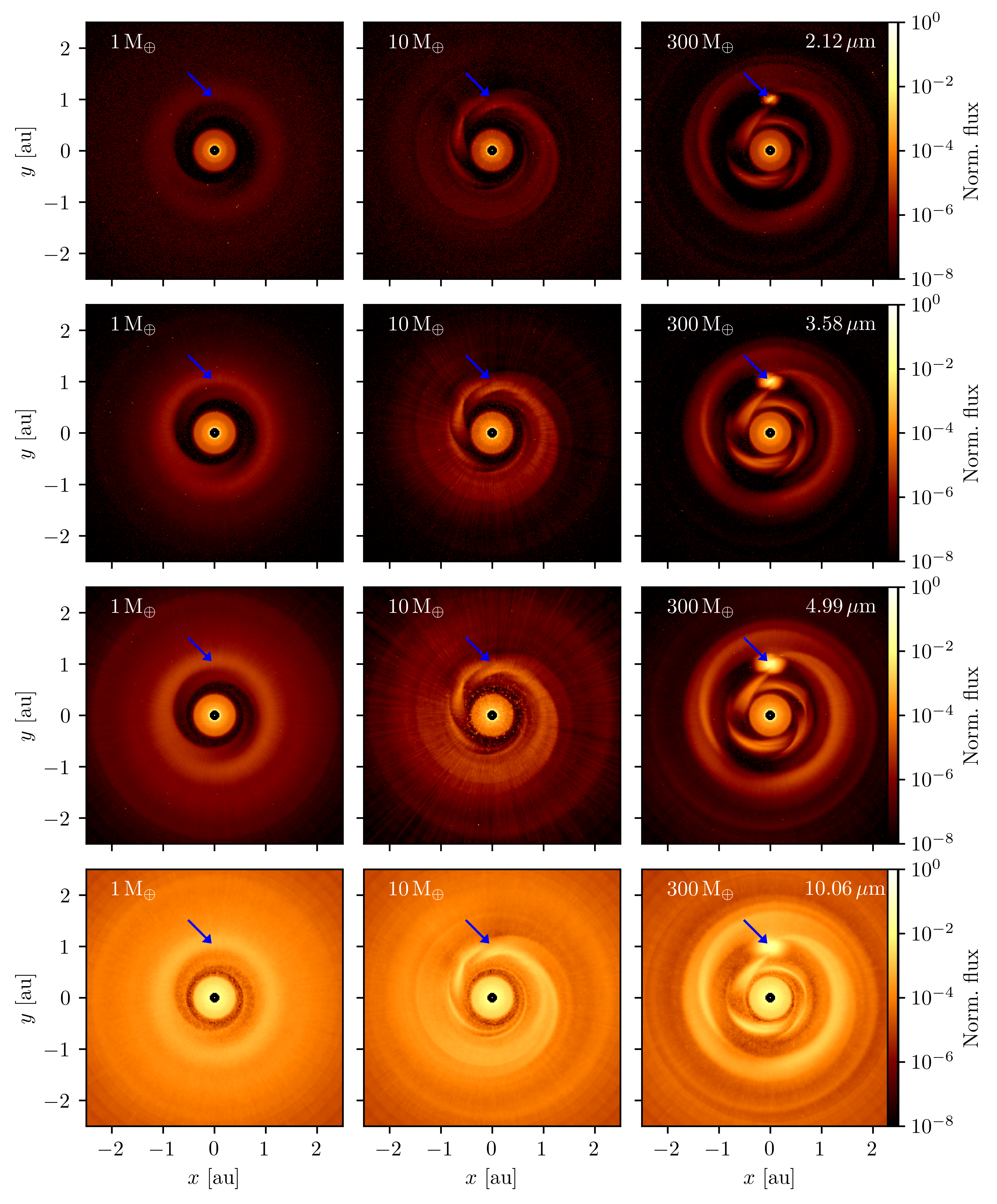}
    \caption{Flux maps for the simulations with planet masses of 1 \ME (left column), 10 \ME (center), and 300 \ME (right) in the K band (first row), L band (second), M band (third), and N band (fourth). The location of the planet, at a radial distance of 1 au north of the center, is indicated with a blue arrow in each simulation. The images are normalized with regard to the flux value of the central pixel where the star is located.} 
    \label{fig:fluxcrossN}
\end{figure*}

Figure \ref{fig:fluxcrossN} shows the simulated flux maps in the K, L, M, and N band for three cases of embedded planets (1, 10, and 300 \ME). The maps are normalized with regard to the maximum flux, that is, the central pixel where the star is located, and show the innermost 2.5 au of the disk. 
While the largest substructures, which are most dominant in the dust densities in Fig. \ref{fig:dustdensities} remain similar for all wavelengths, some details change depending on the specific wavelength band. 
Smaller, fainter ring sections or spiral arms in the N band become even fainter in the other three bands, while the reduced relative flux of some spiral arms increases the contrast of underlying, fainter features, especially in the case of the 300 \ME planet. \\
One particularly important aspect of these simulations is the decrease of the normalized flux of the disk features in the K, M, and L band, compared to those in the N band flux maps. 
In the K, M, and L bands, the brightness of the disk substructures typically ranges between $10^{-7}$ to $10^{-4}$ of the maximum flux, which is a factor of more than $10^2$ fainter than in the N band simulations. 
This is caused by two effects: First, a decrease in absolute stellar flux at the central pixel from the K to the N band, which in the case of the 300 \ME planet decreases from $0.429$ Jy in the K band to $0.0462$ Jy in the N band, and second, an increase in absolute flux in the disk substructures due to the prevalence of colder dust in the disk. 
According to Wien's displacement law, the corresponding temperature of a black body with a spectral radiance peak in the N band at $10.06$ microns is \ca288 K, whereas a maximum in the K band at $2.12$ microns corresponds to a temperature of \ca1372 K.
The change in disk flux between the K band and N band simulations, indicate a larger abundance of dust with temperatures closer to the \ca288 K of the theoretical maximum of the N band, than to the \ca1372 K of the K band. 
Additionally, the optical depth varies over the simulated wavelengths, which enables the radiation of different disk depths to reach the observer and thus causes the flux maps to show different disk layers, which may be of different temperatures and dust densities.
The bright spot at the planetary accretion region in the 300 \ME simulation shows a slower spatial decline in normalized flux in the L band, and thus becomes more prominent compared to the background substructures.

Depending on the mass of the embedded planet, the substructures appear different, with a similar behavior as observed in the dust density in the midplane, that is, a higher normalized flux for a higher planet mass.
For the lowest mass planet of 1 \ME, the simulated N band flux map shows a bright, almost symmetric ring with a radius of \ca 1 au and a width of \ca 0.3 au, which contains slight rotational asymmetries. 
Further inward a bright, hollow disk feature exists between a radius of \ca 0.4 au and the inner disk radius of 0.1 au, with a radial darkening from \ca 3\,\% of the maximum flux at the inner edge, to \ca 0.3\,\% of the maximum at the outer edge. 
This bright inner disk remains similar across all three simulations, both in appearance and brightness magnitude, with only small variations in the 300 \ME simulation.
Additionally, two faint spiral arms are visible, one originating at the bright inner disk in the lower left and the other in the upper right, with both moving clockwise toward the bright ring at \ca1 au orbital distance.
The flux map of the 10 \ME planet features a prominent, non-continuous spiral with a small shadow cast behind the position of the planet, showing the same helicity as the spiral arms in the 1 \ME case.
Before fading out, a split in the spiral arm occurs to the east of the star, creating a two-tailed structure.\\

For a planet mass of 300 \ME, the most significant difference compared to the lower mass scenarios is the luminosity of the planetary accretion region, which becomes the dominant feature with a normalized flux of up to 0.07 per pixel and a size of \ca 0.5 au in diameter.
Apart from this, a spiral emanates radially outside from the gap region, similar to the case of the 10 \ME simulation, but brighter and with a discontinuation after a three-quarter revolution from its innermost point. 
The origin of this spiral is located at the accretion region of the planet, and revolves clockwise outward, similar to the previous spiral. 
The spiral shows a nearly constant radial extent of \ca 0.4 au that increases after one full revolution to \ca 0.5 au.
From the planetary orbit inward, multiple smaller structures exist, most of which are centered around a radial distance of \ca 0.6 au, with one connecting to the accretion region. \\

In order to study the observability of the substructures or their respective influence on the observations, a high relative flux compared to both the star and the immediate disk background is beneficial. 
Thus, the L, M, and N band simulations are the most valuable for further analysis, as their substructures show the highest relative flux and brightness contrast.\\

\begin{figure*}
    \centering
    \includegraphics[width=0.95\linewidth]{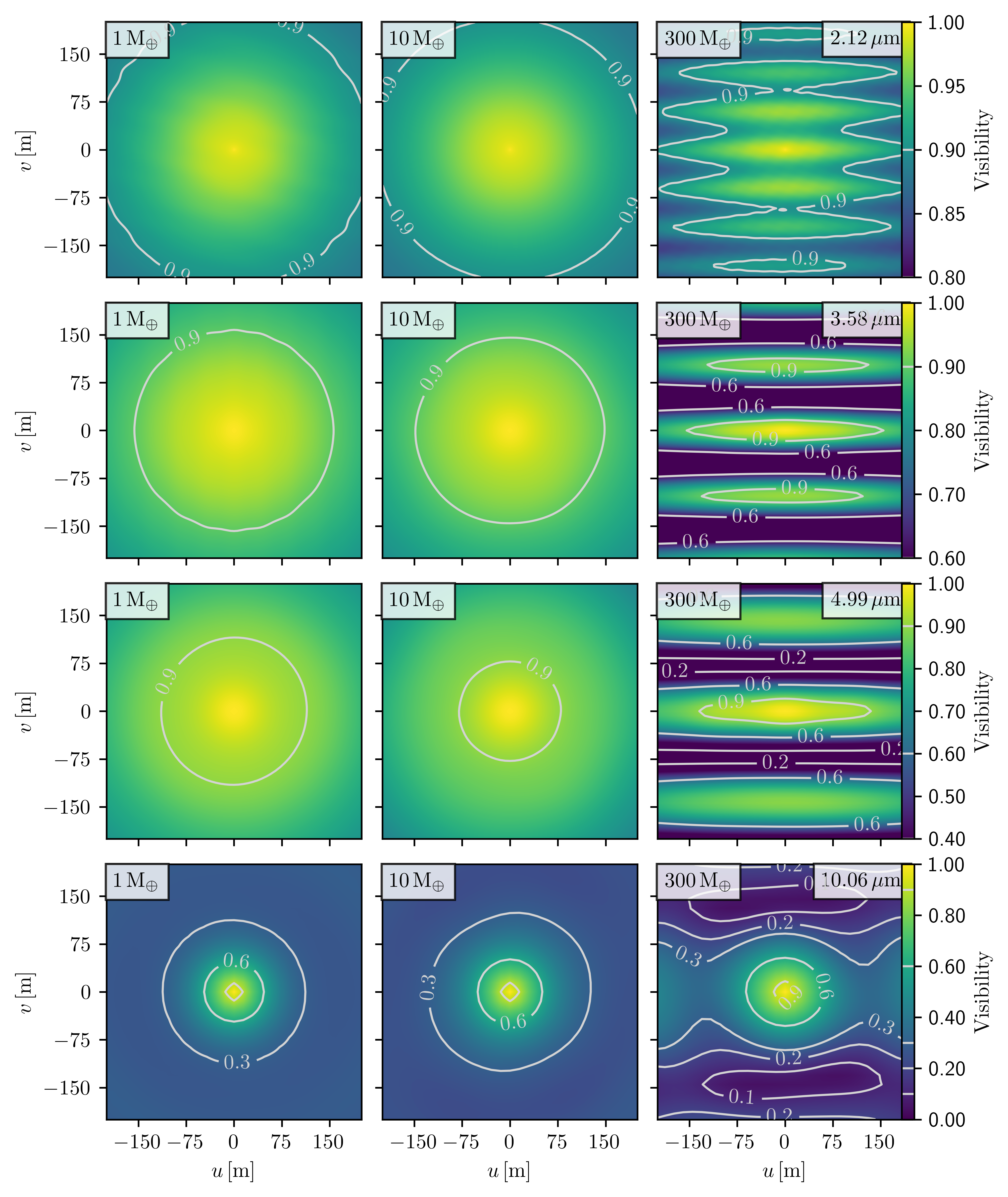}
    \caption{Simulated uv-plane visibilities in the K band (first row), L band (second), M band (third), and N band (fourth)  for the simulations with planet masses of 1 \ME (left column), 10 \ME (center), and 300 \ME (right).  }
    \label{fig:visibilities}
\end{figure*}

\subsection{Simulated visibilities}\label{chap:visibility}
According to the van-Cittert-Zernike theorem (\citealt{cittert1934} and \citealt{zernike1938}), the Fourier transform of an intensity distribution of a distant source is equal to the complex visibilities, which encode the intensity, size, and location of different substructures.
This complex quantity can be measured with interferometers in terms of its absolute value, the visibility, and its angle, the (closure) phase.
A single interferometric observation measures the visibility and the phase at one point in the uv-plane, corresponding to the length and orientation of the projected baseline, that is, the distance between the involved telescopes.
While a single interferometric measurement samples only a single point, multiple observations during a single night measure multiple points along an elliptical path in the uv-plane that depends on the celestial path of the observed object, which directly depends on the declination of that object.
Multi-epoch observations sufficiently spaced in time are able to observe long term changes of the targets system, including the orbital motion of potential planets or surrounding disk material. 
This enables an effectively higher sampling of the uv-plane without changing the location or orientation of the baselines, considering that a rotation of the image plane also causes a rotation in its Fourier transform. 

In the following, we analyze the Fourier transforms in the entire accessible uv-plane of the previously discussed flux maps (Sect. \ref{chap:flux}).
For this purpose we apply a two-dimensional Fourier transform using the Python library \texttt{galario}, chosen for its efficient performance. 
In the next step the obtained results are used to identify approaches to detect and characterize disk substructures observationally.
In particular, the feasibility to constrain potential substructures without relying on the high coverage required for an image reconstruction, will be discussed.
Accurately simulating real VLTI observations would require a number of additional assumptions about observing conditions and target parameters that were outside the focus of this work and were thus not considered.

A key inherent characteristic of the simulated flux maps (Sect. \ref{chap:flux}) is their azimuthal asymmetry or lack thereof, respectively. This property is carried over via the Fourier transform into the two-dimensional visibility distribution. 
Similar to an undisturbed disk, which is showing no substructures (see Fig. \ref{fig:visibility_examples}), the flux map of a 1 \ME planet results in an almost perfectly rotationally symmetric Fourier transform (i.e., visibility distribution) in the uv-plane, albeit with a different radial profile (see Fig. \ref{fig:visibilities}, left column).
Asymmetric planet-induced structures are the most prominent substructures observed in the flux map simulations involving planets with masses of 10 \ME and 300 \ME. Analyzing the characteristic features of these structures in interferometric data may yield valuable insights for identifying similar structures in real observations.

Shown in the center and right column in Fig. \ref{fig:visibilities} are the simulated visibilities of the simulated flux maps for planet masses of 10 \ME and 300 \ME. 
Despite its visibly asymmetric flux map, the visibility pattern in the uv-plane of the 10 \ME planet mass simulation still appears as a nearly rotationally symmetric Gaussian profile, with only a minor deviation visible in the contour lines. 
In contrast, the visibilities in the uv-plane of the 300 \ME planet mass simulation in the right column in Fig. \ref{fig:visibilities} show pronounced, characteristic features, that are absent in an undisturbed disk. 

In the N band (fourth row in Fig. \ref{fig:visibilities}), all three cases show a nearly point-symmetric inner pattern for baselines smaller than \ca 75\,m, with visibilities >0.5 and up to 1 in the center. 
In the 300 \ME case (fourth row on the right in Fig. \ref{fig:visibilities}), the pattern for baselines >100\,m, shows an area of visibilities lower than 0.1 at the top and the bottom, and between 0.3 and 0.5 left and right to the center. Toward longer baselines in both directions of the u-axis, the visibilities increase again. 
This pattern can be explained as consisting of three distinct visibility patterns created by the dominant substructures, overlaid atop each other. 
First, there is a radially decreasing and azimuthally symmetric component mainly resulting from the brightness distribution of the central star and the overall radially decreasing flux of the disk. 
Second, alternating stripes of high and low visibilities, parallel to the u-axis, are introduced by the elevated brightness distribution of the planetary accretion region, which, in combination with the central star, represents a binary structure. 
And lastly, the spiral substructure creates a visibility pattern similar to a Bessel-function, as a thin ring would (see Fig. \ref{fig:visibility_examples}), and is responsible for the \qq{bottlenecks} along the u-axis at $\pm$ 100\,m. 
Although the spiral has a small opening angle, it is not symmetric as a thin ring is, and thus the resulting Bessel-function-like visibility pattern is slightly elongated toward the top left and bottom right.
The influence of each pattern on the cumulative visibility pattern depends on their individual relative intensities.

The right plots in the second and third row of Fig. \ref{fig:visibilities} show the simulated visibilities for the 300 \ME planet mass case, as seen in the L and M band. 
The dominant pattern here are the stripes, which are created by the bright planetary accretion region and the star, which again resemble that of a binary structure.
The central visibility within each stripe decreases in both the u and v directions. In between these stripes, minima reach values typically below 0.6 in the L band and down to 0 in the M band. 
A similar pattern can be seen in the right plot in the first row in Fig. \ref{fig:visibilities} for the 300 \ME planet case in the K band, albeit with a higher overall visibility, where the minima show a value of about 0.85, and thus a smaller gradient than in the L or M band. 
In addition, the aforementioned change of the visibility across the stripes, along the u direction, can be seen more clearly. 
This pattern can be explained by the same dominant binary feature of the accretion region and the star. 
The lower relative flux from the accretion region and the underlying substructures, compared to the star, cause the visibilities to be overall higher than in the N, M, or L band. 
\begin{figure*}
    \centering
    \begin{subfigure}[b]{0.475\textwidth}
         \includegraphics[width=\textwidth]{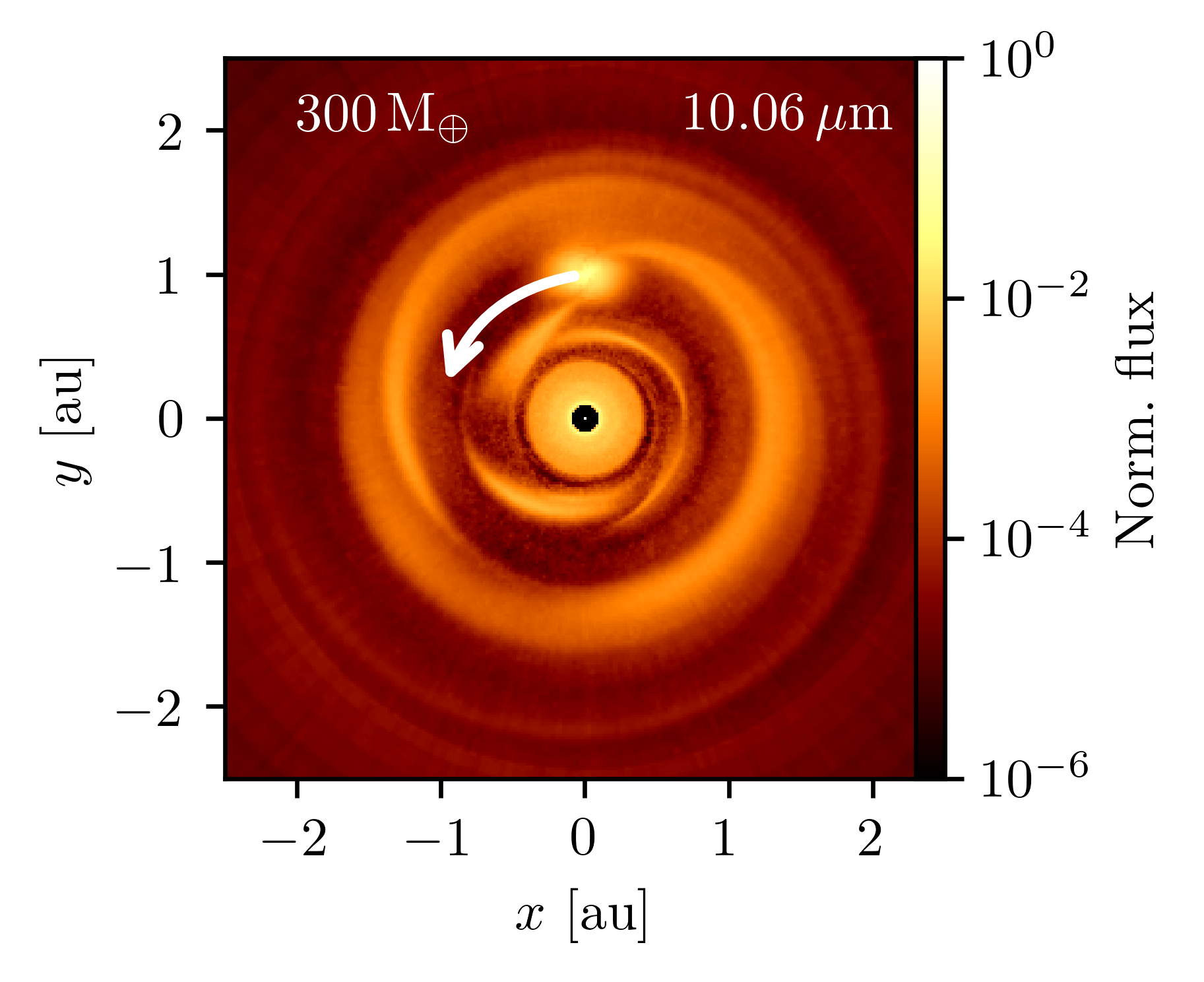}  
         \label{fig:300_n_flux_motion}
    \end{subfigure}
    \hfill
    \begin{subfigure}[b]{0.5\textwidth}
         \includegraphics[width=\textwidth]{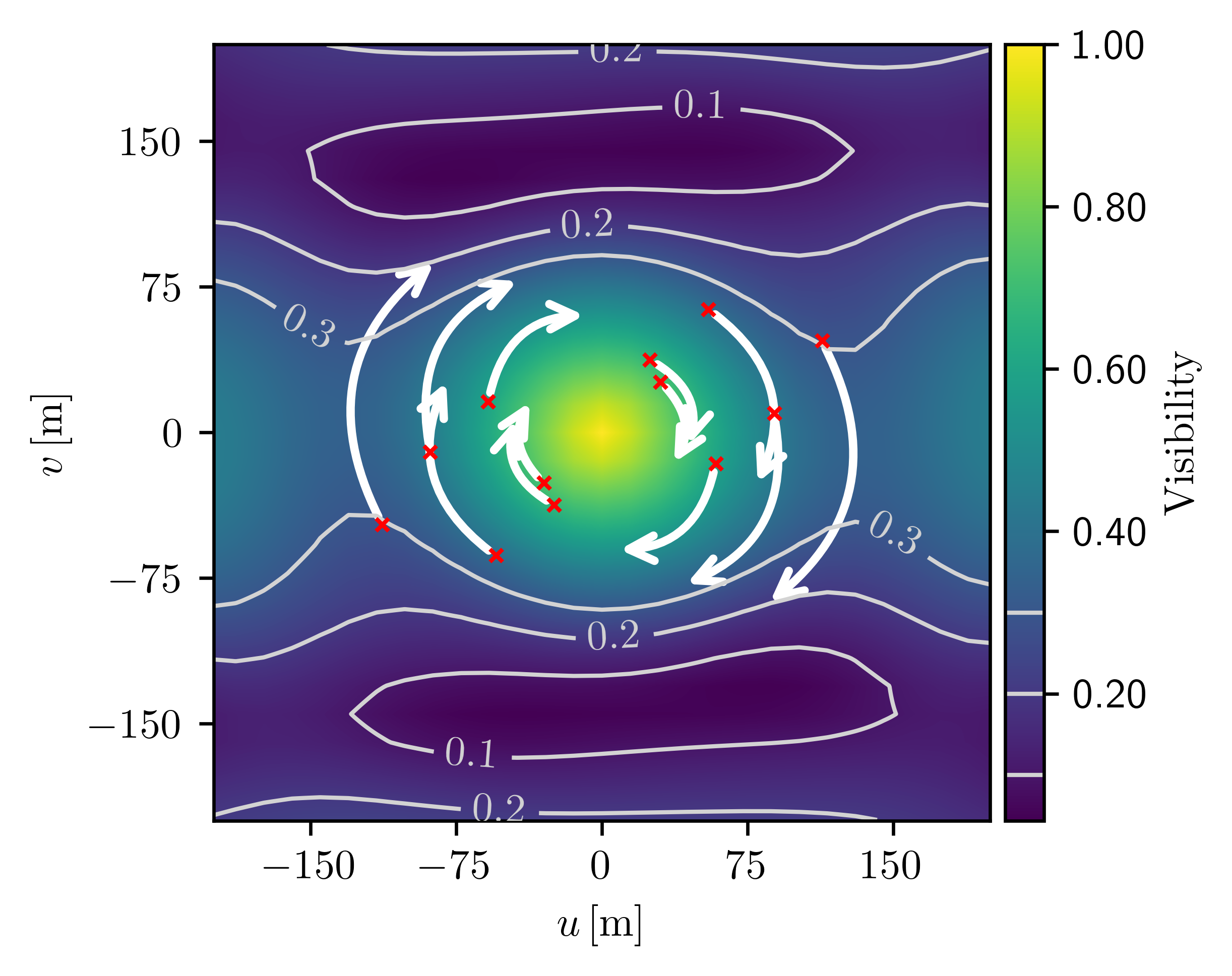} 
         \label{fig:300_n_vis_motion}
    \end{subfigure}
    \caption{Simulated flux map (left) and visibilities in the uv-plane (right) in the N band for a 300 \ME  planet. The red crosses in the uv-plane correspond to the VLTI UT baselines for an observation of an object with a similar declination as HL Tau with its highest possible elevation. The white arrow in the flux map indicates the direction of the orbital motion of the planet and other substructures, while the white arrows in the uv-plane indicate the corresponding, effective change of the UT baselines.} 
    \label{fig:300_motion}
\end{figure*}  

\begin{figure*}
    \centering
    \includegraphics[width=0.75\textwidth]{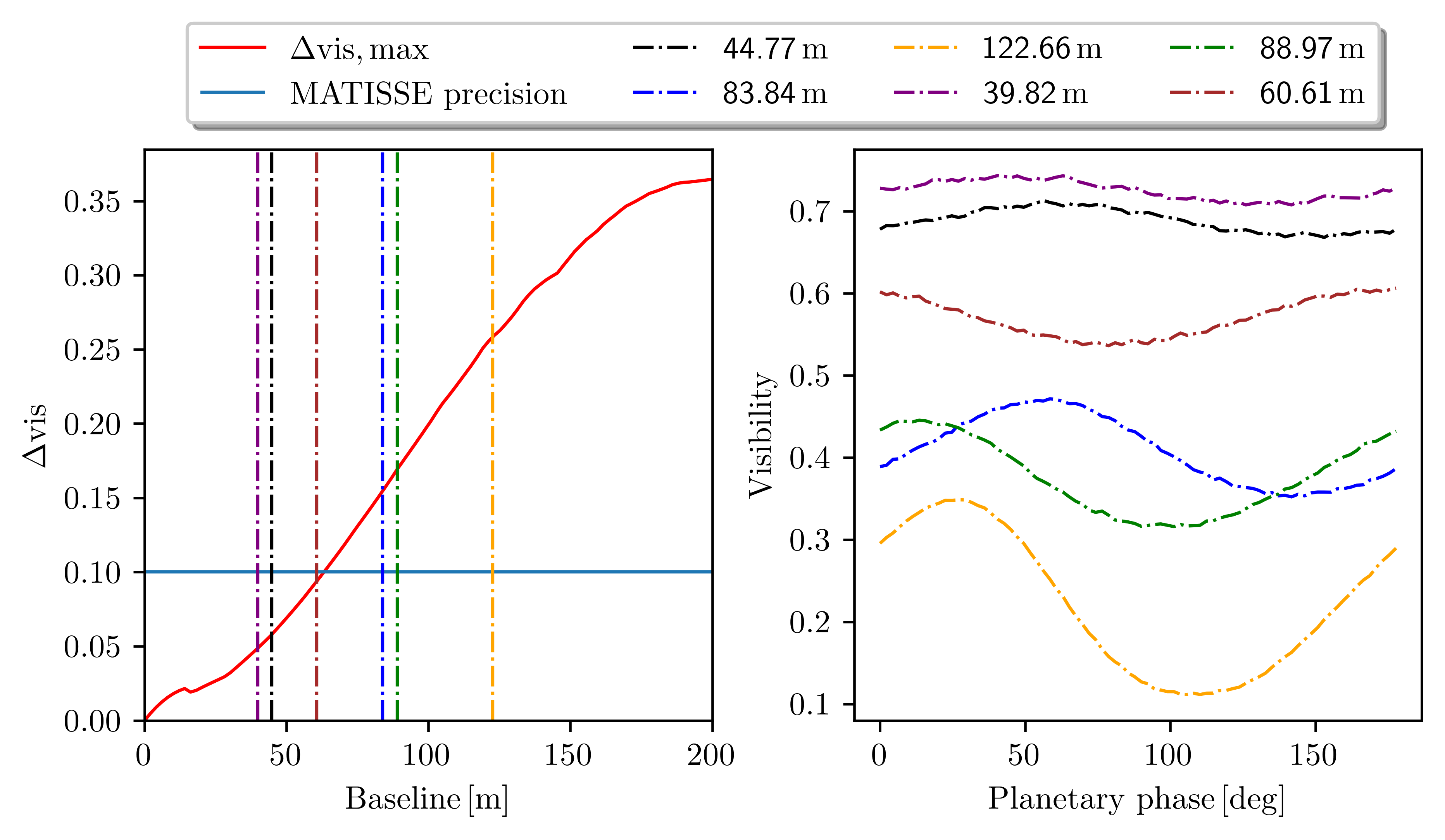} 
    \caption{Interferometric variability in the simulated N band visibilities at radii in the uv-plane corresponding to all UT baselines for an embedded planet of 300 \ME. Shown are the maximum variability, dependent on baseline length (left), and the measured visibility for a given baseline, dependent on the orbital motion of the planet and other disk substructures (right).} 
    \label{fig:varrot300}
\end{figure*}

As mentioned earlier in this chapter, this work places a focus on multi-epoch observations that use the orbital motion of embedded planets and other substructures over time to increase the uv-coverage with static baselines.
Figure \ref{fig:300_motion} shows the effect of multi-epoch observations in the flux map (left) and the uv-plane (right) in the case of the 300 \ME planet in the N band.
The orbital motion of the planet and other disk material in the direction of the white arrow in the flux map causes a rotation of the uv-plane in the same direction, which, for better clarity, is converted to a rotation of the baselines (red crosses) in the opposite direction.
Depending on the orbital motion of the planet and other material, the UT baselines would sample the uv-plane along a circular path and therefore measure a variable visibility (and closure phase).
In Fig. \ref{fig:varrot300} the maximum variability of the measured visibility (i.e., the difference between the highest and lowest measured visibility) is shown as a function of baseline length (left) and angular polar coordinate change (right) due to the orbital motion, assuming the VLTI UT baselines.
The blue, horizontal line represents the detection limit of VLTI/MATISSE of 0.1 in the visibilities (see \citealt{petrov2020} for the MATISSE instrument description), and the vertical lines indicate the different UT baselines. 
Only the variability for the UT1-UT3 (83.84\,m, blue), UT1-UT4 (122.66\,m, yellow), and UT2-UT4 (88.97\,m, green) baselines exceed the guaranteed precision of VLTI observations and are thus detectable.
For smaller baselines (\(<60 \) m), the variance is significantly smaller and will not be distinguishable from the statistical error of the measurements. 
In the right plot, the values of the simulated visibilities in the uv-plane are shown for semicircles at different radii, where each curve corresponds to one UT baseline, emulating the effect of the orbital motion in the image plane. 
These curves represent all possible multi-epoch observations of the target during one full orbit of the embedded planet and allow measuring the angular distance between their extrema, which amounts to \ca90\textdegree \, for most baselines, and  \ca80\textdegree \,for the UT1-UT4 (122.66\,m, yellow) baseline.
This behavior indicates that the maximum change of measured visibilities can be expected for a 90\textdegree \,difference in baseline orientation in the uv-plane, which can be achieved with either multi-epoch observations or, in special cases, single observations at different points during one night, depending on the declination of the object.
A similar behavior is found in the other wavelength bands, albeit with a larger variability in the L and M band, and with a smaller, but for GRAVITY still detectable, variability of less than 0.1 in the K band (see appendices \ref{fig:varrot300_K} to \ref{fig:varrot300_M}).

The approach of using multi-epoch observations to acquire a higher effective sampling in the uv-plane was investigated exclusively for the case of face-on oriented disks in this work.
An inclined disk with an orbiting circumplanetary accretion region and other substructures would also create visibility patterns.
These would change depending on their corresponding orbital position and brightness, which can be diminished by vertically extended disk features in the case of high inclination angles.
The aforementioned multi-epoch observation approach would therefore still be valid for these inclined disks.
However, an inclined disk would resemble an elliptical shape on the sky, and thus create an additional rotationally asymmetric visibility pattern, which has to be considered when calculating the variability in the visibilities between observations.
Additionally, the apparent separation of a potential bright accretion region, as can be seen for the 300 \ME planet case, from the star would change during an orbit.
This would influence the measured visibility by changing the visibility pattern in the uv-plane.
\subsection{Closure phases}\label{chap:closphase}
The other observable of interferometric observations is the phase of the complex visibilities. 
In order to eliminate atmospheric noise, a common practice consists of combining the phases obtained by a triplet of telescopes and calculating the so called closure-phase (see, e.g., \citealt{monnier2007}).
The closure phases can provide spatial information of the brightness distribution of an observed object, where a closure phase of 0\textdegree \,indicates a rotationally symmetric flux source, while closure phases $\neq$0\textdegree \,indicate asymmetric, off-center flux components.

\begin{figure*}
    \centering
    \includegraphics[width=0.75\linewidth]{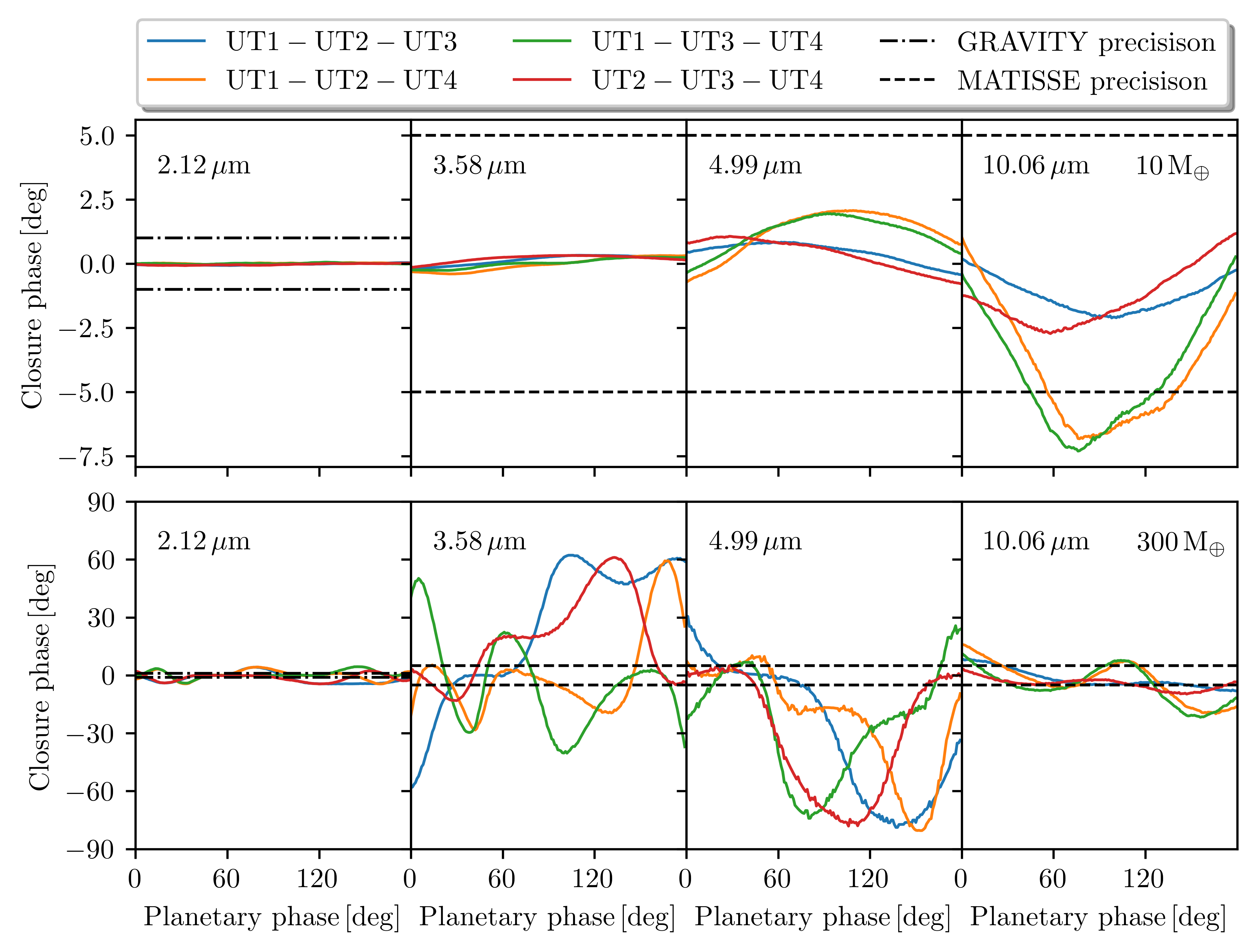}
    \caption{Simulated closure phases of all baseline triplets during half a planetary orbit for the simulations with a planet mass of 10 \ME (top) and 300 \ME (bottom) in the K band (first column), L band (second), M band (third), and N band (fourth). The precision of VLTI/MATISSE is shown as the dashed black lines at $\pm$ 5\textdegree in the L, M, and N band, while the precision of GRAVITY in the K band is shown as the dashed, dotted black line at $\pm$ 1\textdegree.}
    \label{fig:closurephases_all}
\end{figure*}
In Fig. \ref{fig:closurephases_all} the simulated closure phases for all baseline triplets in the simulations for the 10 \ME and 300 \ME (top and bottom rows) planet mass cases are shown in the K band, L band, M band, and N band (columns from left to right), assuming a counter clockwise rotation of the baselines of up to 180\textdegree\,, representing the orbit of the substructures between multi-epoch observations. 
The direction of this rotation is chosen arbitrarily in this case, but the principle works in the opposite direction as well.

For K band observations, the GRAVITY instrument achieves a precision of 1\textdegree\, in the closure phases \citep{gravity2017}, which is beyond the simulated values in the 1 and 10 \ME case, but is exceeded in the 300 \ME case, up to \ca 5\textdegree\,, indicating a detectable non-zero signal.
The instrument VLTI/MATISSE achieves a precision of 5\textdegree\, in the closure phases (see \citealt{petrov2020}), which is above the maximal values of the simulated closure phases for all wavelengths considered for the 1\ME planet case (not shown here), as well as for the 10 \ME planet in the L and M band, which indicates no detectable asymmetry in the brightness distribution of these disks. 
The N band flux maps (see Fig. \ref{fig:fluxcrossN}) show a higher, though still weak, contrast of the substructures against the background, which prevails as a closure phase signal in the single digits, only surpassing the 5\textdegree\, limit for two baseline triplets for the 10 \ME planet. 
Thus, despite the absence of any potentially detectable traces of the substructures in the visibilities, their influence on the closure phases could be traced. 

In strong contrast to these rather small closure phases, for the 300 \ME planet, the simulated closure phases in the N band (Fig. \ref{fig:closurephases_all}, bottom right) range between \ca -20\textdegree\, and \ca 10\textdegree\, for two triplets, and between \ca -10\textdegree \, and \ca 10\textdegree \, for the other two. 
Even larger closure phases are predicted for the L and M band (bottom center), where every triplet reaches in absolute terms values of up to \ca50\textdegree\, indicating a significant and detectable influence of asymmetric flux sources. 
Since the flux maps in the L and M band show a dominant planetary accretion region for the 300 \ME planet, it can be considered the main cause of the measurable closure phases, and thus may provide an opportunity to determine the flux ratio between the accretion region and the star from the amplitude of the closure phases.

All simulations with detectable closure phases further show a change of sign for sufficiently spaced observations, as the substructures orbit around the star.
A time dependent change in the sign of the closure phases is therefore an indicator of an orbiting structure and can be used to differentiate from non-orbiting off-center sources with variable flux.

\section{Discussion}\label{chap:discussion}
Based on our findings, we discuss the morphology of the substructures in the flux maps, and compare them with those observed at larger scales (Sect. \ref{chap:submorph}), give an estimation for the observability of the visibility variability using the VLTI/GRAVITY and VLTI/MATISSE (Sect. \ref{chap:observ}), provide a modeling approach based on multi-epoch observations (Sect. \ref{chap:modeling}), and discuss multiple ways to differentiate from other sources of variability (Sect. \ref{chap:differentiation}).

\subsection{Substructure morphology}\label{chap:submorph}
A principal goal of this study was to investigate whether substructures at smaller spatial scales were similar to those at larger spatial scales, where embedded planets could potentially be responsible for their existence.
We find that the simulated flux maps in this work show similar substructures compared to those observed at larger scales \citep{bae2023}, and show a morphological dependence on the planet mass, and an observational dependency on the wavelength.

A ring and gap structure, similar to the substructure in the flux map of the 1 \ME planet (see Fig. \ref{fig:fluxcrossN}), has been observed multiple times and ranks as the most common substructure type, according to \citet{bae2023}. 
This class of substructure was observed in the case of the prominent example of HL Tau \citep{alma2015}, with multiple rings and gaps as far out as 97 au. 
A number of mechanisms have been proposed to explain the origin of rings and gaps apart from the possible influence of embedded planets (see, e.g., \citealt{van_der_Marel_2019}). 
Our work shows an existing ring for the smallest planet mass flux map, indicating that these planets are a plausible cause for rings even at small scales, and should motivate further research into these structure formation mechanisms.
The substructures in this case appear centro-symmetric in the closure phases (less than 0.5\textdegree\, across all considered wavelengths), and lack a detectable trace in the visibilities (see Fig. \ref{fig:visibilities}), due to a low contrast ratio against their background.

In the simulations with the two higher planet masses of 10 \ME and 300 \ME, the dominant substructure is a spiral. 
\citet{bae2023} mentions a lower occurrence of spirals than rings, with an abundance of observations at higher disk-to-star mass ratios in the targets, although this could potentially be due to an observational bias. 
\citet{Yu_2019} found a strong correlation between the pitch angle, that is, the opening width of the spiral, and the disk mass, where a higher disk mass tends to correlate with a smaller pitch angle. 
Because the disk masses in this work were equal among all simulations, no investigation into this relationship was possible. 
For our constant disk mass it was found that I) observable spirals only appear for planet masses larger than 10 \ME, II) a higher planet mass leads to a more pronounced spiral against the disk background, and III) increasing the planet mass decreases the pitch angle.

The flux maps in the case of planets with masses of 10 \ME and 300 \ME (see Fig. \ref{fig:fluxcrossN}) additionally show a variety of substructures in the inner 1 au of the disk, which resemble neither rings nor spirals and are best described as ring sections. 
These features appear radially thinner and azimuthally smaller than the other, more dominant substructures, while being only slightly fainter than them. 
The closest analogue of substructures at larger scales are crescents, based on their azimuthal expanse and relative thickness.
Contrary to the crescents observed at larger scales (e.g., HD142527 in \citealt{Boehler_2018} or HD 143006 in \citealt{andrews2018}) which are parts of a ring with an increased brightness, the features here are separate entities without an underlying ring.
Some of these features are radially tilted, that is one end is closer to the center than the other. 
This could imply their nature as azimuthally limited, uncoiled spiral arms.

In the considered parameter space, an increase of the planet mass results in increasingly prominent, brighter substructures (see Fig. \ref{fig:fluxcrossN}). 
The substructures are brightest in the simulated normalized N band flux maps, while the K, L, and M band simulations show them by a factor of up to 1000 fainter.
Only the planetary accretion region in the 300 \ME planet mass simulation remains bright in the L and M band, and even improves its contrast to the background substructures, compared to the K band.

\subsection{Observability of planet-induced structures}\label{chap:observ}
Of the visible substructures in the simulated flux maps, only that of the 300 \ME planet shows a visibility pattern in the uv-plane for baselines <200\,m that differs noticeably from the visibility pattern of an unperturbed disk.
The planet-induced structures of the other planet masses, that is, the faint ring and gap in the flux map of the 1 \ME planet, and the disturbed spiral caused by the 10 \ME planet, show no detectable traces in the simulated visibilities, and can thus not be differentiated from an otherwise undisturbed disk.
Multi-epoch observations with the unmovable UTs and with sufficient temporal distance (days, months, or years apart) may utilize the orbital motion of the embedded planet, to sample the uv-plane at different polar angles, which would otherwise require a change in baseline orientation on the ground.
With this increased sampling of the uv-plane, multiple measurements of the visibility for each baseline produce a variability (i.e., the difference between the lowest and highest detectable visibility).
The feasibility of this approach relies on the interferometric variability in the visibilities to be stronger than the instrument sensitivity of GRAVITY in the K band of 0.01 and of 0.1 for MATISSE in the L, M, and N band. 
In case of the 300 \ME planet, the highest possibly measurable variability in the visibilities with the UT baselines is 0.24 in the N band, 0.93 in the M band, 0.52 in the L band, and 0.058 in the K band, which is detectable by both GRAVITY and MATISSE.

The timescale between the extrema of measurable visibility depends on the orbital motion of the embedded planet and other substructures.
For example, a planet at 1 au on a Keplerian orbit around a pre-main-sequence star of mass 0.5 \(\text{M}_{\odot}\) has an orbital period of 512 days, corresponding to an angular velocity of 0.69\textdegree\, per day. 
The angular distance of the local extrema of the visibility amounts to about 90\textdegree (see Fig. \ref{fig:varrot300}), equal to a quarter orbit, or in this example a time span of 128 days. 
The exact value depends on the influence of smaller asymmetric structures that can cause a deviation. 
As can be shown on the basis of simple toy-models, flux maps with multiple different asymmetric structures (archimedean spirals, ring sections, and secondary Gauss profiles) and varying parameters for these (relative flux, size, opening angle, radial length, FWHM), are expected to show a broad range of possible angular distances for the maximum variability. 
The deviation from 90\textdegree\, particularly depends on the relative flux of the asymmetric structure compared to that of the planetary accretion region. 

Considering the closure phases, the VLTI/MATISSE achieves a precision of 5\textdegree\, (see \citealt{petrov2020}), while VLTI/GRAVITY achieves a precision of 1\textdegree\, in the K band \citep{gravity2017}. 
These limits are exceeded for multi-epoch closure phase measurements, in case of the 10 \ME planet case only in the N band, whereas for the 300 \ME planet case they are exceeded across all wavelengths considered (see Fig. \ref{fig:closurephases_all}).
Despite the absence of traces of corresponding substructures in the visibilities, the Fourier transform of the 10 \ME planet case shows detectable closure phases of up to -7\textdegree\, in the N band, while the clear asymmetric features in the M band flux map of the 300 \ME planet case produce closure phases of up to -80\textdegree\, (see Fig. \ref{fig:closurephases_all}). 
Similar to the visibilities, the orbital motion of the planet leads to a detectable variability in the closure phases, and a change in sign for most baseline triplets. 
The 1 \ME planet case shows no detectable closure phases for any of the considered wavelengths, which is consistent with the absence of strong asymmetric features in the simulated flux map.\\

In summary, only the 300 \ME planet causes observable variability in both the visibilities, as well as the closure phases, while the 10 \ME planet has a sufficiently significant impact only on the variability of the closure phases.
The 1 \ME planet causes no interferometrically detectable signatures.
Thus, observing both the visibilities and closure phases allows for a coarse estimation of the planetary mass, via the presence or absence of variability in these observables. 
A finer sampling of planet masses between the current cases, and a more detailed knowledge of the substructures and their influence on the visibilities and closure phases would, however, be necessary to enable finer estimations. 
A dominant feature in the flux maps, that is, the bright planetary accretion region, is present in the case of the 300 \ME planet and absent in the case of the 1 and 10 \ME planets (see Fig. \ref{fig:fluxcrossN}). 
This feature strongly influences both, the visibility and closure phases, that can be used as a relatively crude observable indicator of the planetary mass, where the lack of variability provides an upper limit of the planet mass. 

Our findings motivate a finer sampling of planetary masses between the currently considered cases of 10 and 300 \ME. 
In particular, this approach could reveal the planetary mass at which the accretion region becomes bright enough to be detected, or when it dominates the otherwise existing substructures. 
Considering general substructures, a better understanding of when a ring breaks up into a spiral (possibly between 1 and 10 \ME), when ring sections increasingly appear (exist at 10 \ME, number increases for higher mass) or whether these ring sections could play a dominating role, compared to spirals (possibly outside the current mass range), would improve our ability to infer the mass of the embedded planet from observations.
The effect of the planetary mass on the observability of substructures is just one of many different influences, however, that would also need to be considered in the interpretation of the presence or absence of observable substructures.
For an otherwise already defined parameter space of the disk age, distance, and photometric variability, for example, this additional constraint on planet mass could still prove advantageous.
However, both increased sampling of planetary masses and deeper investigation into other influences on the observability of substructures are beyond the scope of the current study.

\subsection{Multi-wavelength analysis approach}\label{chap:modeling}
The variability and characteristic visibility pattern caused by the planetary accretion region of the 300 \ME planet may not be sufficient to deduce the existence of an embedded planet. 
In particular, it is beneficial to search for underlying substructures, caused by the induced density waves of an embedded planet. 

When comparing the visibilities for the 300 \ME case in Fig. \ref{fig:visibilities} (right column), one finds that all of them are dominated by the visibility pattern caused by the accretion region.
Only in the N band (fourth row, right column), the spiral structure results in a slightly elliptical feature. 
This can be described by differently stretched zero-order Bessel functions in u and v direction, with the first minimum with almost vanishing visibility at baselines of \ca 150 m. 
However, this influence is too faint to accurately be induced from single observations. 
In the following, we evaluate the potential of using the simultaneous observations in the L and N bands with VLTI/MATISSE.

The simulated flux map for the 300 \ME planet case shows a significant wavelength dependence, as seen in the flux maps in Fig. \ref{fig:fluxcrossN} (right column), where the distinctive visibility patterns can be attributed to the change in the relative flux of the various disk features. 
In the N band flux map for the 300 \ME planet, the underlying substructures, that is, the spiral at 1 au and the inner ring at 0.1 to 0.4 au, are significantly brighter than in the other considered wavelength bands, which is mentioned in more detail in Section \ref{chap:flux}, likely due to a higher abundance of colder dust and optical depth effects. 
While the planetary accretion region reaches a maximum relative flux of almost 10\,\% of the star at this wavelength, these underlying substructures are bright enough to influence the visibility pattern.
The already mentioned binary source-like visibility pattern (see Fig. \ref{fig:visibilities}) that is caused by radiation of the star and the planetary accretion region results in a local maximum of the visibilities for theoretical baselines of 300\,m in the v-direction, which lies outside the baseline lengths accessible with the VLTI. 
For shorter wavelengths, such as 3.58\,$\mu$m in the L band in Fig. \ref{fig:visibilities} (fourth row, middle column), this local maximum moves to shorter baselines and can be traced at baselines of \ca 80\,m. 
Additionally, the relative flux of the planetary accretion region decreases less than the relative flux of the underlying substructures, and causes the accretion region to dominate the visibilities in the uv-plane as a result. 

Both of these aspects combined, result in a distinctly observable pattern in the visibilities from which the accretion region of the embedded planet can be approximated as a Gaussian-profile shaped source with a certain central intensity, width and separation from the star. 
However, it requires a sufficient uv-coverage to identify both the visibility pattern and its first maximum.
Coupled with the closure phases, it is possible to create a simple model with both the separation, flux ratio, and orientation (via the direction of the visibility pattern) of the accretion region of the embedded planet. 

The L band flux map in Fig. \ref{fig:fluxcrossN} shows no other significant influences in the L band visibilities for this 300 \ME embedded planet, which would have to be differentiated from this binary signal. 
While the M band flux map in the same figure also features a prominent planetary accretion region that predominantly shapes its visibilities, the background disk structures are brighter in the M band than in the L band, which is therefor preferred.

The next step requires a fitting process between the observed visibilities and the visibilities corresponding to the modeled flux map. 
While this fit will result in fitted parameters for the location, flux ratio, and width of the planetary accretion region, only the location parameters can be considered identical across all wavelengths.
The flux ratio and width of the accretion region, on the contrary, will remain specific to the L band, and have to be refitted for the other wavelengths.
For the N band, the binary source fit has to be repeated, but now the location parameters can be set from the previous fit.
After achieving a sufficiently high quality in this fitting process, the binary signal can be subtracted from the measured visibilities.
Apart from potential residual flux values at the stellar and planetary accretion region, the resulting visibility pattern will then only consist of the underlying substructures.  
Subsequent models of rings or spirals can then be fitted to the resulting visibilities and may enable the study of underlying substructures.
This approach thus combines the clear, dominant signal of the accretion region in the L band, with the higher relative flux of the substructures in the N band. 
Comparing these substructures with those expected from hydrodynamic simulations for a planet with a given mass and orbital distance, may provide stronger arguments for the presence or absence of an embedded planet in the disk.

\subsection{Source differentiation}\label{chap:differentiation}
The orbiting, asymmetric substructures and the bright environment of the embedded planet, lead to a detectable variability in the observed visibilities and closure phases, which must be differentiated from other phenomena with similar observational signatures (see, e.g., \citealt{kospal2012}). 

Spatially unresolved observations \citep[e.g.,][]{borgesfernandes2012,2025arXiv250207900S, 2019MNRAS.485.1614O}, have already allowed one to differentiate between multiple sources of temporal variations originating in changes of the stellar flux (e.g., cool spots, hot spots, or `dippers'), while interferometric observations enable the investigation of the contribution of asymmetric structures. 
Objects with a known photometric variability are therefore the preferred targets for a study on the possible interferometric variability.
Such a study on interferometric variability can be found in \citet{kobus2020}, where the main differences in the observations lie in the amplitude of the variability of the squared visibilities ($\Delta_{vis^{2}}$) and the shortest timescale of detectable visibility differences ($\Delta t_{\text{var}}$), the results of which are compiled in table \ref{tab:var_sources} together with our own results.
Additionally, \citet{kobus2020} presented a method to differentiate the physical origin of two variability types, depending on whether the change in visibilities between observations is uniform or irregular, that is described in the following paragraphs.

A classification into symmetric and asymmetric variability sources is possible depending on whether the visibilities obtained at all baselines either increase or decrease between observations, or partially increase and partially decrease, respectively. 
This behavior originates from the visibility patterns that contain the rotational symmetry or asymmetry of the disk. 
For a symmetric Fourier transform, each baseline will experience no change in visibilities under a rotation around the center, which would show in the case of multi-epoch observations. 
Any change in visibility due to a variable stellar flux will change the visibilities uniformly and lead to an increase or decrease along all baselines (in the case of a face-on disk). 

Since the VLTI UT baselines are not aligned in a straight line but have multiple orientations in the uv-plane (visible in Fig. \ref{fig:300_motion} as red crosses), an asymmetric pattern may cause some baselines to measure an increased visibility during a revolution of the substructures around their host star, while others may show a visibility decrease.
This is shown on the right of Fig. \ref{fig:varrot300}, where the change of the visibilities for all UT baseline under one half orbit of the embedded planet is simulated for the N band of the 300 \ME case. 
Due to their different orientation on the ground, the change in visibilities is non-uniform. 
For example, two observations performed at planetary phases of 0 and 60 degrees (which correspond to a minimum separation of 85.3 days for this system) would show a decrease of the visibilities for the baseline UT3-UT4 (60.61m, brown line) and an increase in visibility for UT1-UT3 (83.84m, blue line), while UT2-UT3 (39.82m, purple line) would remain almost constant.
Analyzing this behavior enables a general classification into symmetric and asymmetric disks based on potentially only two observations. 
These observations would need to be sufficiently spaced in time to achieve the required coverage in the uv-plane. 

As a note of caution, it has to be considered that it is possible for a rotationally asymmetric disk to appear symmetric with this approach, should all baselines sample locally symmetric parts of the uv-plane. 
In the case of the 300 \ME planet, the visibilities in the uv plane show no detectable variability for baselines $<60\,$m, which correspond to larger structures in the disk (see the left of Fig. \ref{fig:varrot300}).

\begin{table}
      \caption{Observed interferometric variability and observational signature of the considered case of the embedded 300 \ME planet.}
    \centering
    \begin{tabular}{c|c|c|c|c}
       Object  & Band & \(\Delta_{\text{vis}^2}\) & \(\Delta t_{\text{vis}}\)  & Mechanism \\
       \hline &  &  &  &  \\
       HD 50138  & K & 0.318& 56d & morphological\\
       & & & & disk change\footnotemark[1]\\
       &  &  &  &  \\
       DX Cha  & H & 0.281 &<1d & eclipsing binary\footnotemark[2] \\
       &  &  &  &  \\
       AK Sco & H & 0.539 & <22d & eclipsing binary\footnotemark[3]\\
       &  &  &  &  \\
       HD 142527  & H & -- &<0.06d & low mass  \\
       & & & & stellar companion\footnotemark[4]\\
       &  &  &  &  \\
       V856 Sco & H & 0.156 & <1d & circumstellar\\
       &  K & 0.531 & <1d &extinction\footnotemark[5] \\
       &  &  &  &  \\
       HD 163296 & H & 0.124 & 12d & dust clouds\footnotemark[6]\\
       &  K & 0.51& <6d & \\
       &  &  &  &  \\
       R Cra& K & 0.464 & <1d  & disk material\footnotemark[7]\\
       \hline &  &  &  &  \\
       300 \(\text{M}_{\Earth}\)& K & 0.11& 6d  & embedded planet\footnotemark[8]\\
       &  L & 0.7& 9d&\\
       &  M & 0.9& 6d&\\
       &  N & 0.1 & 29d& \\
       
    \end{tabular}
    \tablefoot{ The values of \(\Delta_{\text{vis}^2}\) and \(\Delta t_{\text{vis}}\) of the literature studies above have been determined by \citet{kobus2020}.}
    \tablebib{(1)~\citet{kluska2016}; (2) \citet{boehm2004}; (3) \citet{andersen1989}; (4) \citet{close2014}; (5) \citet{perez1992}; (6) \citet{rich2019}; (7) \citet{sissa2019}; (8) This work.}
    \label{tab:var_sources}
\end{table}

\citet{kobus2020} further differentiates sources of interferometric variability between morphological disk changes for HD 50138 \citep{kluska2016}, the spectroscopic binaries DX Cha \citep{boehm2004} and AK Sco \citep{andersen1989}, a low-mass stellar companion of HD 142527 at a distance of 12 au from the star \citep{close2014}, circumstellar extinction in the case of V856 Sco \citep{perez1992}, dust clouds in the system of HD 163296 \citep{rich2019}, and obscuration by disk material of R Cra \citep{sissa2019}.
Table \ref{tab:var_sources} gives an overview of these objects, with their corresponding variability amplitude in the squared visibilities and shortest timescale of significant variability, as well as their established mechanism that causes the variability.

The only sensible comparison between the simulated systems in this work, and those covered by \citet{kobus2020} is the 300 \ME planet mass case in the K band.
This is because the simulations for the lower planet masses show no detectable variability, while the other wavelengths are not directly comparable to H and K band observations.
For the K band observation from \citet{kobus2020}, the variability amplitude lies between 0.318 for HD 50138 and 0.531 for V856 Sco, while the K band simulation from this work shows a smaller amplitude of only 0.11 in the squared visibilities. 
The values for all wavelengths from this work are calculated as the largest difference between the maximum and minimum detectable squared visibility per baseline.

Considering the observed timescales of detectable variability in the K band, V856 Sco and R Cra showed a variability timescale of 1 day, which was the shortest covered timescale by their respective observations. 
For the morphological changes of HD 50138, the timescale was 56 days, while the dust clouds of HD 163296 showed a timescale of less than 6 days in the K band. 
The variability timescale in this work is defined as the time between two observations that show a variability of more than 0.1 in the squared visibilities in the L, M, and N band, which is the precision of the VLTI/MATISSE instrument, and more than 0.01 in the K band visibilities, corresponding to the precision of the VLTI/GRAVITY instrument.
Here, the K band variability timescale was estimated to be around 6 days, which comes close to the observed timescale of HD 163296, but is noticeably shorter than the 56 days of HD 50831.
The similarity with HD 163296 does not continue in the visibility amplitude, which deviates from the simulation of this work by a factor of five.

In order to asses the uniqueness of the individual or combined quantities of the other wavelengths, further interferometric observations of systems with known mechanisms of variability are required. 
Both the L and M band show a remarkably large variability amplitude, while for the N band the longest timescale of the simulated disks in this work of 29 days was found, which could potentially be used as a criterion to identify embedded planets.

The aforementioned possibility of a change in the angular distance of maximal variability due to substructures, is another way of creating a timescale closer to non-planetary sources. 
This influence is small for the cases discussed in this work, with a deviation of less than 10\textdegree\, for the N band visibilities of the 300 \ME planet.
An inclined disk, or an embedded planet with an eccentric orbit, which is, however, less likely due to predicted effects of eccentricity dampening (see, e.g., \citealt{cresswell2008}), could also change the angular distance and thus produce a variability timescale that is much closer to other sources and complicate the differentiation.

\section{Conclusions and outlook}
Based on hydrodynamic and subsequent radiative transfer simulations of close-in protoplanets, embedded in an optically thick protoplanetary disk, we investigated the feasibility to detect characteristic disk substructures resulting from planet-disk interactions at sub au scales.
We find that long-baseline interferometry with the VLTI/GRAVITY and VLTI/MATISSE will allow tracing these structures, assuming a distance of the targets of 140\,pc, corresponding to the rich, low-mass star-forming regions in Taurus-Aurigae.
Compared to the K, L, and M band, in the N band observations, the substructures are brighter by a factor of $10^3$ compared to the expected stellar brightness. 
These structures split into symmetric and asymmetric structures, with the latter being dependent on the planetary mass and changing from symmetric rings to asymmetric spirals for an increasing planet mass. 
The morphology of the simulated structures in the K, L, M, and N band is consistent with those observed at larger scales, except for crescents that are noticeably absent here. 
The embedded planets in our derived flux maps are located, depending on their mass, at either the inner rim of a ring feature (lower mass) or the beginning of a spiral arm (higher mass).\\
Solely the brightness distribution for the case of a 300 \ME planet shows an increased flux in its planetary accretion region, which is a dominant feature in the simulated flux maps across all wavelengths and strongly influences the visibility pattern.
Other underlying structures only noticeably affect the N band visibilities, while the L and M band visibilities are mainly shaped by the planetary accretion region.
Using simultaneous multi-wavelength observations may allow us to exploit this behavior and enable the investigation of the fainter structures from the N band visibilities, after removing the influence of the accretion region on the visibilities, derived from simulated L or M band observations. 

The asymmetric structures visible in the 300 \ME case are detectable in both the visibilities and closure phases. 
The closure phases in the case of the 10 \ME planet also show a detectable asymmetric flux, but no detectable impact on the visibility.
The considered 1 \ME planet does not have any observable impact on the visibility and closure phase.

The visibility pattern of the 300 \ME planet simulation depends on the orientation of the baselines, and thus varies with changing baseline orientations. 
A 90\textdegree\, change in baseline orientation maximizes this effect and can be achieved by observing the target in multiple epochs, after it completes a quarter orbit. 
The behavior of a non-uniform change of visibilities over all observed baselines, as presented by \citet{kobus2020}, can also be found in the simulations originating from asymmetric disk substructures in this work and is a possible indicator for these.

A source differentiation is possible via the unique combination of the variability timescale and amplitude across multiple bands, where the disk brightness distribution in the case of a 300 \ME planet shows a low variability amplitude of 0.11 and 0.1 in the K and N band, with a significant variability after 6 days in the K band, and 29 days in the N band. 
The simulated L and M band observations show a large variability amplitude of 0.7 and 0.9 in the squared visibilities with a variability timescale of 9 days and 6 days respectively, a combination not seen in any of the other discussed physical phenomena potentially causing variability.
To summarize, both multi-epoch and multi-wavelength interferometric observations with high-resolution instruments, such as VLTI/MATISSE or VLTI/GRAVITY, have the potential to trace signatures and thus the physics of a protoplanet-disk interaction on sub au to au scales.

\begin{acknowledgements}
We acknowledge the support of the DFG priority program
SPP 1992 "Exploring the Diversity of Extrasolar Planets" under contracts WO 857/17-2.
\end{acknowledgements}

\bibliography{References}

\begin{appendix}

\onecolumn
\section{Visibility examples}
\begin{figure*}[!ht]
    \centering
    \includegraphics[width=0.7\linewidth]{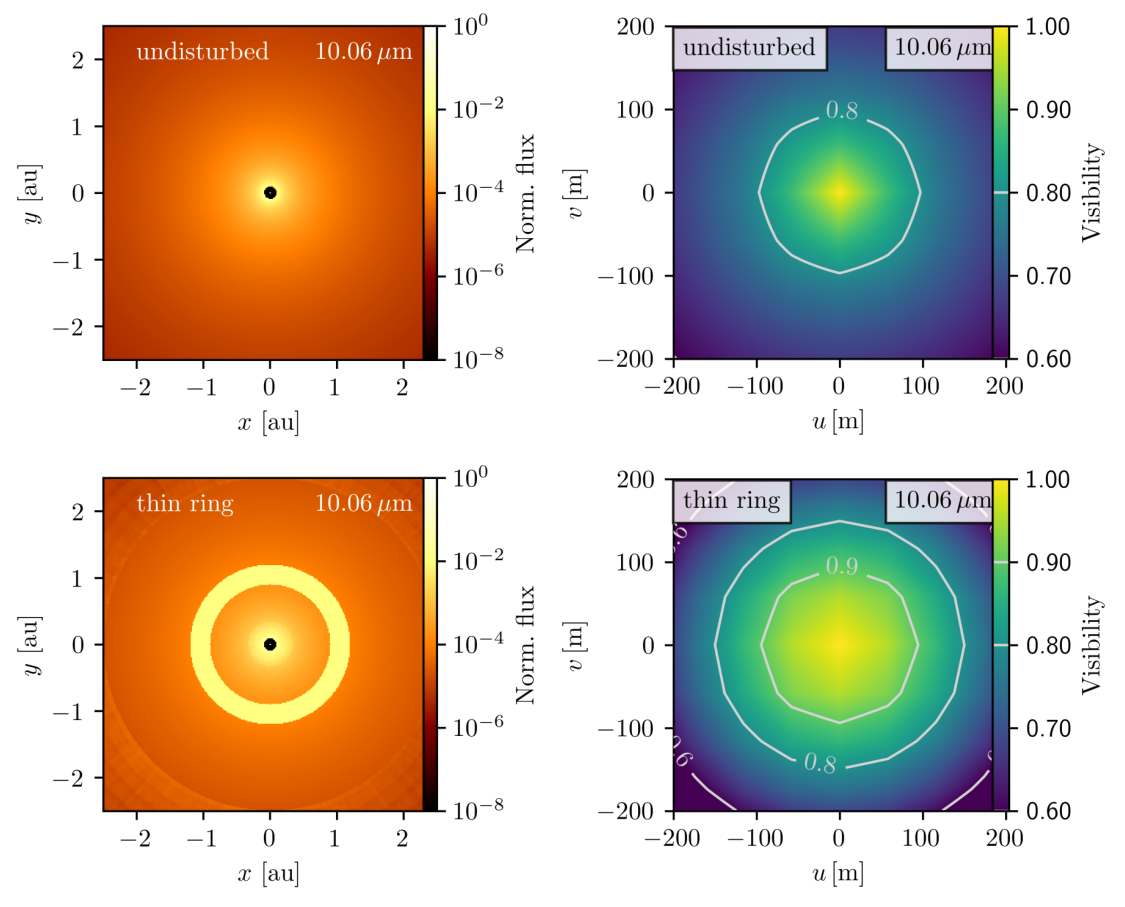}
    \caption{N band flux maps (left) and corresponding visibilities in the uv-plane (right) for examplary cases of an undisturbed disk (top row) and a thin ring (bottom row), respectively.}
    \label{fig:visibility_examples}
\end{figure*}

\section{Variability in K, L, and M bands for the case of an embedded 300 \ME planet}\label{appx:variability}

The following figures show the variability of the simulated visibilities for an embedded planet of 300 \ME in different wavelength bands. Shown is the maximum variability as a function of baseline length (left), and the measured visibility for a given UT baseline, dependent on the orientation in the uv-plane (right). 

\begin{figure*}[!ht]
    \centering
    \includegraphics[width=0.7\textwidth]{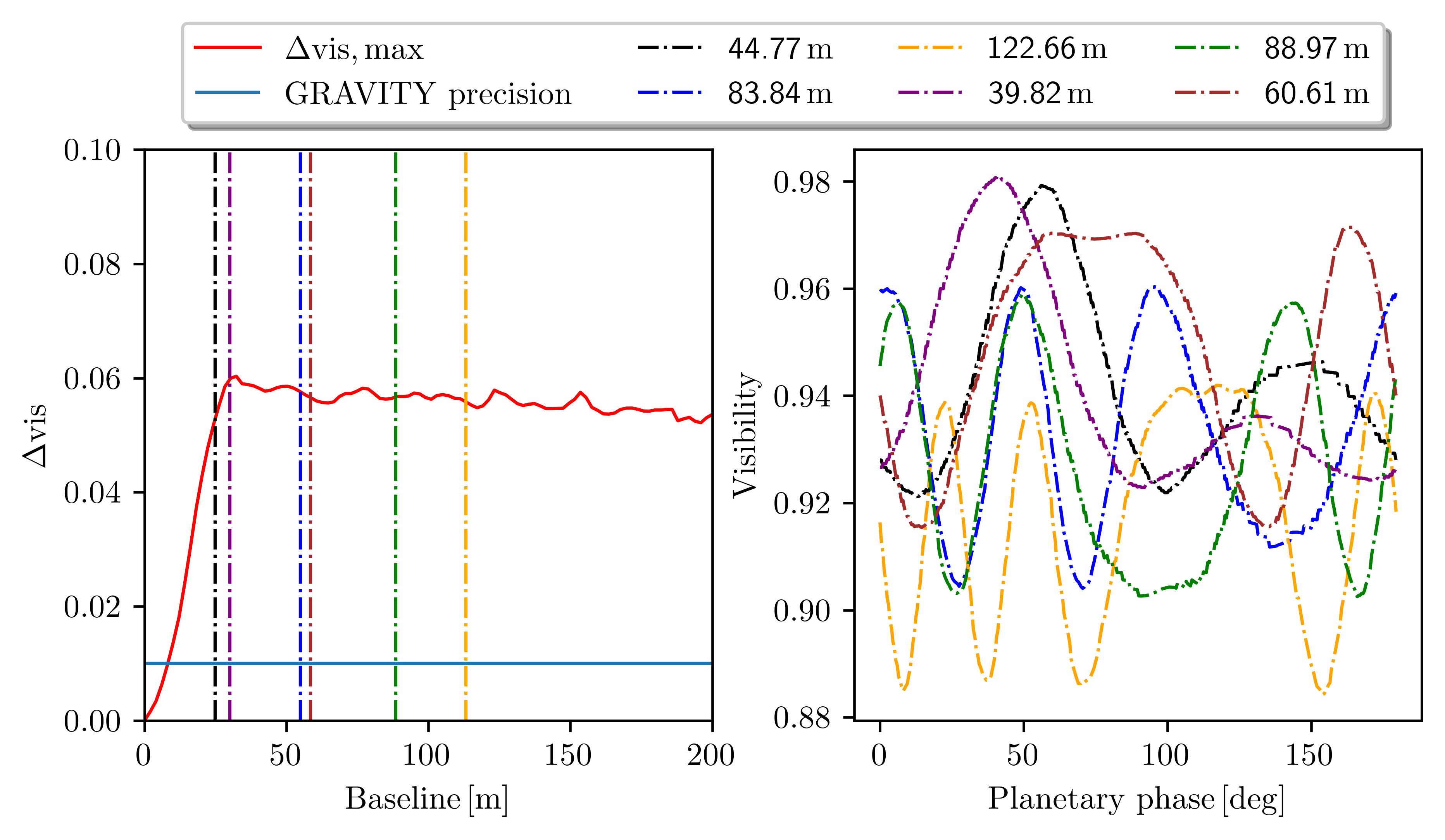} 
    \caption{Variability in the K band.} 
    \label{fig:varrot300_K}
\end{figure*}

\begin{figure*}[!ht]
    \centering
    \includegraphics[width=0.7\textwidth]{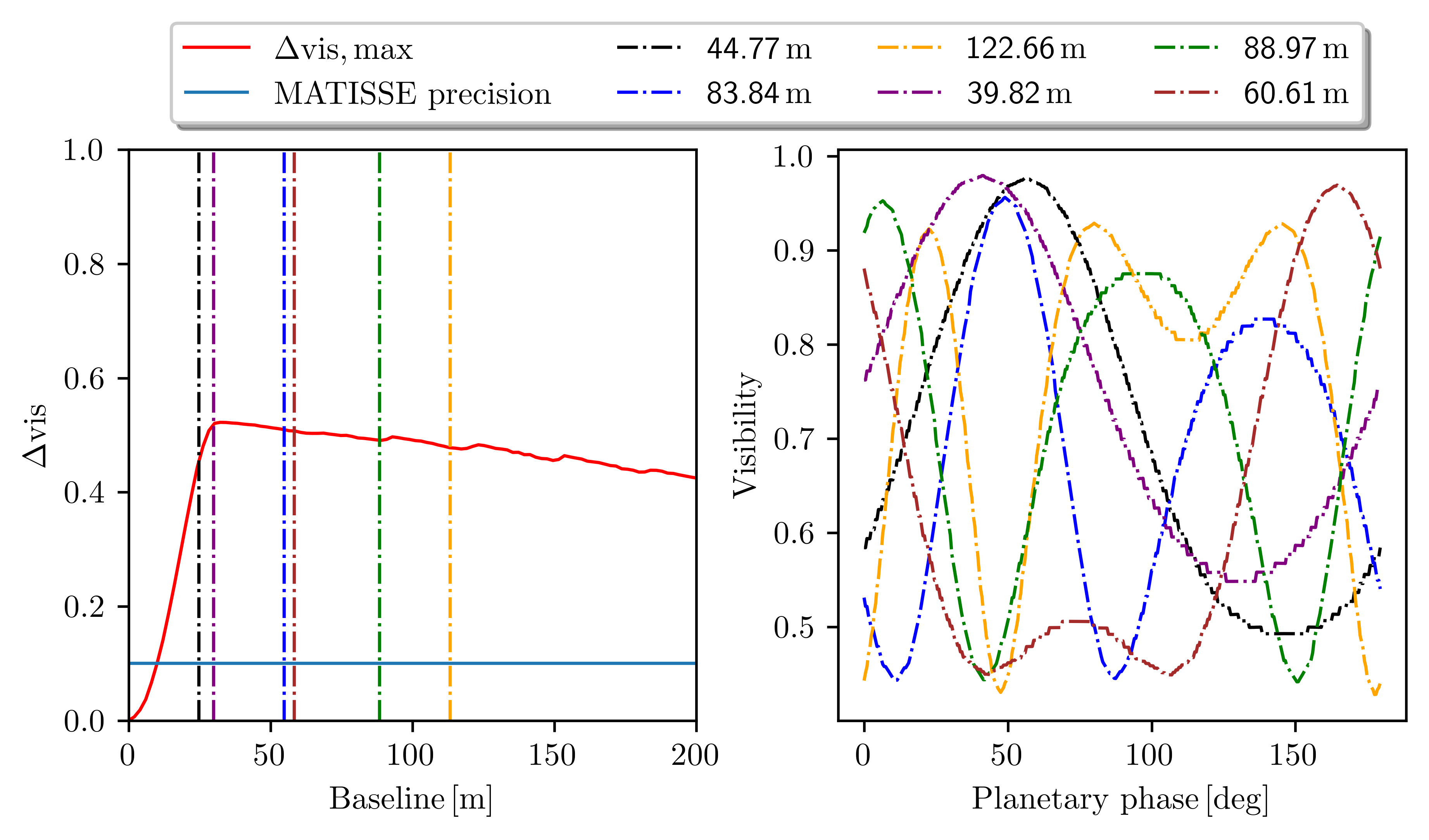} 
    \caption{Variability in the L band.} 
    \label{fig:varrot300_L}
\end{figure*}

\begin{figure*}[!ht]
    \centering
    \includegraphics[width=0.7\textwidth]{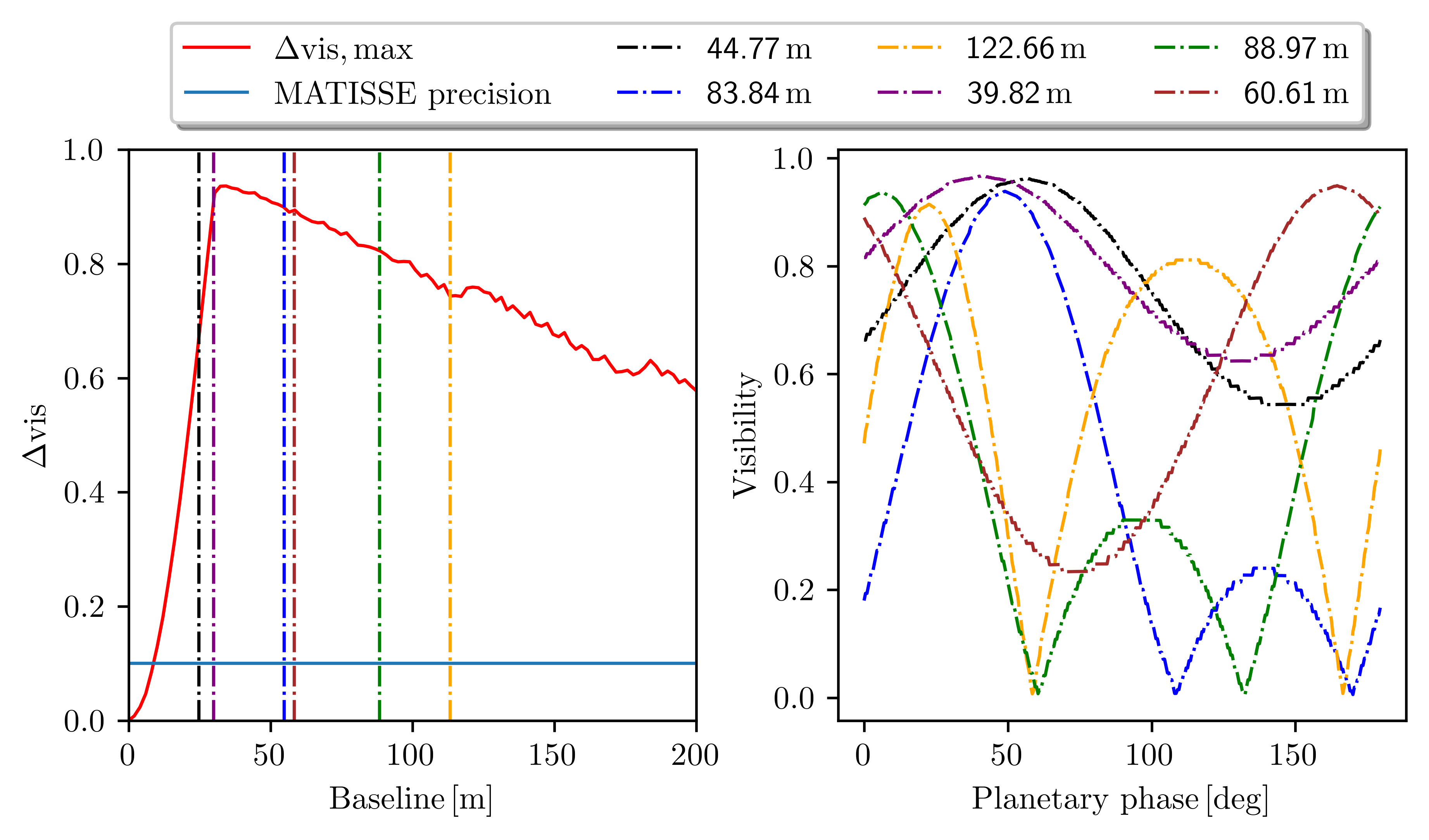} 
    \caption{Variability in the M band.} 
    \label{fig:varrot300_M}
\end{figure*}

\end{appendix}

\end{document}